\documentclass[twocolappendix,appendixfloats,]{emulateapj}

\usepackage[colorlinks = true, linkcolor = blue, urlcolor  = blue, citecolor = blue, anchorcolor = blue]{hyperref}
\usepackage{hyperref}
\usepackage{apjfonts}
\usepackage{rotating}
\usepackage{amsmath}
\usepackage{color}
\usepackage[dvipsnames]{xcolor}
\usepackage{CJK}

\newcommand{\thetae}{\theta_{\rm E}}
\newcommand{\pie}{\pi_{\rm E}}

\definecolor{darkbrown}{RGB}{139,69,19}

\shorttitle{OGLE-2018-BLG-1011L\lowercase{b,c}: Microlensing Multiplanetary System}
\shortauthors{Han et al.}

\begin{document}

\title{OGLE-2018-BLG-1011L\lowercase{b,c}: Microlensing Planetary System with 
Two Giant Planets Orbiting a Low-mass Star}

\author{
Cheongho~Han\altaffilmark{001}, David~P.~Bennett\altaffilmark{002,003,1001},
Andrzej~Udalski\altaffilmark{004,1002}, Andrew~Gould\altaffilmark{005,006,007,1003},
Ian~A.~Bond\altaffilmark{008,1001}, Yossi~Shvartzvald\altaffilmark{009,1003,1004},
Kay-Sebastian Nikolaus\altaffilmark{010}, Markus~Hundertmark\altaffilmark{010}, 
Valerio~Bozza\altaffilmark{011,012}, Arnaud~Cassan\altaffilmark{013},
Yuki~Hirao\altaffilmark{014,1001}, Etienne Bachelet\altaffilmark{015}\\
(Leading authors),\\
and \\
Michael~D.~Albrow\altaffilmark{016}, Sun-Ju~Chung\altaffilmark{005,017},  
Kyeongsoo~Hong\altaffilmark{001},
Kyu-Ha~Hwang\altaffilmark{005}, Chung-Uk~Lee\altaffilmark{005}, 
Yoon-Hyun~Ryu\altaffilmark{005}, In-Gu~Shin\altaffilmark{005},  
Jennifer~C.~Yee\altaffilmark{018}, Youn~Kil~Jung\altaffilmark{005}, 
Sang-Mok~Cha\altaffilmark{005,019}, Doeon~Kim\altaffilmark{001},
Dong-Jin~Kim\altaffilmark{005}, 
Hyoun-Woo~Kim\altaffilmark{005}, Seung-Lee~Kim\altaffilmark{005,017}, 
Dong-Joo~Lee\altaffilmark{005}, Yongseok~Lee\altaffilmark{005,019}, 
Byeong-Gon~Park\altaffilmark{005,017}, Richard~W.~Pogge\altaffilmark{006} \\
(The KMTNet Collaboration),\\
Przemek~Mr{\'o}z\altaffilmark{004}, Micha{\l}~K.~Szyma{\'n}ski\altaffilmark{004},
Jan~Skowron\altaffilmark{004}, Radek~Poleski\altaffilmark{006},
Igor~Soszy{\'n}ski\altaffilmark{004}, Pawe{\l}~Pietrukowicz\altaffilmark{004},
Szymon~Koz{\l}owski\altaffilmark{004}, Krzysztof~Ulaczyk\altaffilmark{020},
Krzysztof~A.~Rybicki\altaffilmark{004}, Patryk~Iwanek\altaffilmark{004},
Marcin~Wrona\altaffilmark{004}\\
(The OGLE Collaboration) \\   
Fumio~Abe\altaffilmark{021}, Richard Barry\altaffilmark{002},           
Aparna~Bhattacharya\altaffilmark{002,003}, Martin~Donachie\altaffilmark{022}, 
Akihiko~Fukui\altaffilmark{023},  Yoshitaka~Itow\altaffilmark{021}, 
Kohei~Kawasaki\altaffilmark{014}, Iona~Kondo\altaffilmark{014}, 
Naoki~Koshimoto\altaffilmark{024,025}, Man~Cheung~Alex~Li\altaffilmark{022},    
Yutaka~Matsubara\altaffilmark{021}, Yasushi~Muraki\altaffilmark{021}, 
Shota~Miyazaki\altaffilmark{014}, Masayuki~Nagakane\altaffilmark{014}, 
Cl\'ement~Ranc\altaffilmark{002}, Nicholas~J.~Rattenbury\altaffilmark{022}, 
Haruno~Suematsu\altaffilmark{014}, Denis~J.~Sullivan\altaffilmark{026}, 
Takahiro~Sumi\altaffilmark{014},  Daisuke~Suzuki\altaffilmark{027}, 
Paul~J.~Tristram\altaffilmark{028}, Atsunori~Yonehara\altaffilmark{029}\\                    
(The MOA Collaboration),\\ 
Pascal Fouqu\'e\altaffilmark{030,031}, Shude Mao\altaffilmark{032,033,034},
Tianshu Wang\altaffilmark{032}, Weicheng Zang\altaffilmark{032}, Wei Zhu\altaffilmark{035},
Matthew~T.~Penny\altaffilmark{006,1004} \\
(The CFHT Collaboration),\\
Charles~A.~Beichman\altaffilmark{036}, Geoffery~Bryden\altaffilmark{037}, 
Sebastiano~Calchi~Novati\altaffilmark{036}, B.~Scott Gaudi\altaffilmark{006}, 
Calen~B.~Henderson\altaffilmark{036}, Savannah~Jacklin\altaffilmark{038}, 
Keivan~G.~Stassun\altaffilmark{038}\\
(The UKIRT Microlensing Team),\\
}

\email{cheongho@astroph.chungbuk.ac.kr}


\altaffiltext{001}{Department of Physics, Chungbuk National University, Cheongju 28644, Republic of Korea} 
\altaffiltext{002}{Code 667, NASA Goddard Space Flight Center, Greenbelt, MD 20771, USA}
\altaffiltext{003}{Department of Astronomy, University of Maryland, College Park, MD 20742, USA}
\altaffiltext{004}{Warsaw University Observatory, Al.~Ujazdowskie 4, 00-478 Warszawa, Poland} 
\altaffiltext{005}{Korea Astronomy and Space Science Institute, Daejon 34055, Republic of Korea} 
\altaffiltext{006}{Department of Astronomy, Ohio State University, 140 W.\ 18th Ave., Columbus, OH 43210, USA} 
\altaffiltext{007}{Max Planck Institute for Astronomy, K\"onigstuhl 17, D-69117 Heidelberg, Germany} 
\altaffiltext{008}{Institute of Natural and Mathematical Sciences, Massey University, Auckland 0745, New Zealand}
\altaffiltext{009}{IPAC, Mail Code 100-22, Caltech, 1200 E.\ California Blvd., Pasadena, CA 91125, USA}
\altaffiltext{010}{Zentrum f\"{u}r Astronomie der Universit\"{a}t Heidelberg, Astronomisches Rechen-Institut, M\"{o}nchhofstr.~12-14, 69120 Heidelberg, Germany}
\altaffiltext{011}{Dipartimento di Fisica ``E.~R.~Caianiello'', Universit\'e di Salerno, Via Giovanni Paolo II, I-84084 Fisciano (SA), Italy}
\altaffiltext{012}{Istituto Nazionale di Fisica Nucleare, Sezione di Napoli, Via Cintia, I-80126 Napoli, Italy}
\altaffiltext{013}{Institut d'Astrophysique de Paris, Sorbonne Universit\'e, CNRS, UMR 7095, 98 bis bd Arago, 75014 Paris, France}
\altaffiltext{014}{Department of Earth and Space Science, Graduate School of Science, Osaka University, Toyonaka, Osaka 560-0043, Japan}
\altaffiltext{015}{Las Cumbres Observatory, 6740 Cortona Drive, Suite 102, Goleta, CA 93117 USA}
\altaffiltext{016}{University of Canterbury, Department of Physics and Astronomy, Private Bag 4800, Christchurch 8020, New Zealand} 
\altaffiltext{017}{University of Science and Technology (UST), 217 Gajeong-ro Yuseong-gu, Daejeon 34113, Republic of  Korea}
\altaffiltext{018}{Center for Astrophysics | Harvard \& Smithsonian, 60 Garden St., Cambridge, MA 02138, USA} 
\altaffiltext{019}{School of Space Research, Kyung Hee University, Yongin, Kyeonggi 17104, Korea} 
\altaffiltext{020}{Department of Physics, University of Warwick, Gibbet Hill Road, Coventry, CV4 7AL, UK} 
\altaffiltext{021}{Institute for Space-Earth Environmental Research, Nagoya University, Nagoya 464-8601, Japan}
\altaffiltext{022}{Department of Physics, University of Auckland, Private Bag 92019, Auckland, New Zealand}
\altaffiltext{023}{Okayama Astrophysical Observatory, National Astronomical Observatory of Japan, 3037-5 Honjo, Kamogata, Asakuchi, Okayama 719-0232, Japan}
\altaffiltext{024}{Department of Astronomy, Graduate School of Science, The University of Tokyo, 7-3-1 Hongo, Bunkyo-ku, Tokyo 113-0033, Japan}
\altaffiltext{025}{National Astronomical Observatory of Japan, 2-21-1 Osawa, Mitaka, Tokyo 181-8588, Japan}
\altaffiltext{026}{School of Chemical and Physical Sciences, Victoria University, Wellington, New Zealand}
\altaffiltext{027}{Institute of Space and Astronautical Science, Japan Aerospace Exploration Agency, 3-1-1 Yoshinodai, Chuo, Sagamihara, Kanagawa, 252-5210, Japan}
\altaffiltext{028}{University of Canterbury Mt.\ John Observatory, P.O. Box 56, Lake Tekapo 8770, New Zealand}
\altaffiltext{029}{Department of Physics, Faculty of Science, Kyoto Sangyo University, 603-8555 Kyoto, Japan}
\altaffiltext{030}{CFHT Corporation, 65-1238 Mamalahoa Hwy, Kamuela, Hawaii 96743, USA}
\altaffiltext{031}{Universit\'e de Toulouse, UPS-OMP, IRAP, Toulouse, France}
\altaffiltext{032}{Physics Department and Tsinghua Centre for Astrophysics, Tsinghua University, Beijing 100084, China}
\altaffiltext{033}{National Astronomical Observatories, Chinese Academy of Sciences, A20 Datun Rd., Chaoyang District, Beijing 100012, China}
\altaffiltext{034}{Jodrell Bank Centre for Astrophysics, Alan Turing Building, University of Manchester, Manchester M13 9PL, UK}
\altaffiltext{035}{Canadian Institute for Theoretical Astrophysics, University of Toronto, 60 St George Street, Toronto, ON M5S 3H8, Canada}
\altaffiltext{036}{IPAC, Mail Code 100-22, Caltech, 1200 E. California Blvd., Pasadena, CA 91125, USA}
\altaffiltext{037}{Jet Propulsion Laboratory, California Institute of Technology, 4800 Oak Grove Drive, Pasadena, CA 91109, USA}
\altaffiltext{038}{Vanderbilt University, Department of Physics \& Astronomy, Nashville, TN 37235, USA}
\altaffiltext{1001}{MOA Collaboration.}
\altaffiltext{1002}{OGLE Collaboration.}
\altaffiltext{1003}{KMTNet Collaboration.}
\altaffiltext{1004}{UKIRT Microlensing Team.}

\begin{abstract}
We report a multiplanetary system found from the analysis of microlensing event 
OGLE-2018-BLG-1011, for which the light curve exhibits a double-bump anomaly around 
the peak.  We find that the anomaly cannot be fully explained by the binary-lens or 
binary-source interpretations and its description requires the introduction of an 
additional lens component.  The 3L1S (3 lens components and a single source) modeling 
yields three sets of solutions, in which one set of solutions indicates that the lens 
is a planetary system in a binary, while the other two sets imply that the lens is a 
multiplanetary system.  By investigating the fits of the individual models to the 
detailed light curve structure, we find that the multiple-planet solution with 
planet-to-host mass ratios  $\sim 9.5\times 10^{-3}$ and $\sim 15\times 10^{-3}$ are 
favored over the other solutions.  From the Bayesian analysis, we find that the lens 
is composed of two planets with masses $1.8^{+3..4}_{-1.1}~M_{\rm J}$ and 
$2.8^{+5.1}_{-1.7}~M_{\rm J}$ around a host with a mass $0.18^{+0.33}_{-0.10}~M_\odot$ 
and located at a distance $7.1^{+1.1}_{-1.5}~{\rm  kpc}$.  The estimated distance indicates 
that the lens is the farthest system among the known multiplanetary systems.  
The projected planet-host 
separations are $a_{\perp,2}=1.8^{+2.1}_{-1.5}~{\rm au}$ ($0.8^{+0.9}_{-0.6}~{\rm au}$) 
and $a_{\perp,3}=0.8^{+0.9}_{-0.6}~{\rm au}$, where the values of $a_{\perp,2}$ in and 
out the parenthesis are the separations corresponding to the two degenerate solutions, 
indicating that both planets are located beyond the snow line of the host, as with the 
other four multiplanetary systems previously found by microlensing. 
\end{abstract}


\section{Introduction}\label{sec:one}

Detecting planetary systems with multiple planets located around and beyond 
the snow line is important for the investigation of the planet formation scenario. 
In planetary systems, snow line $a_{\rm snow}$ indicates the distance from the central 
star at which the temperature is low enough  for volatile compounds such as water, methane, 
carbon dioxide, and carbon monoxide to condense into solid ice grains. 
According to the standard theory of planet formation, i.e., core-accretion theory 
\citep{Mizuno1980, Stevenson1982, Pollack1996}, giant planets are thought to form around 
the snow line because the solid grains of water and other compounds rapidly accumulate 
into large planetary cores that will eventually grow into giant planets.  The high efficiency 
of giant planet formation in this region may help the formation of multiple planets as 
demonstrated by the existence of the two giant planets in the Solar system, i.e., Jupiter 
and Saturn.  Furthermore, there can be multiple snow lines corresponding to the individual 
compounds of protostar nebula and each snow line may be related to the formation of specific 
kinds of planets \citep{Cleeves2016}.  Around a Sun-like star, for example, the water snow 
line would roughly correspond to the orbit of Jupiter and the carbon monoxide snow line would 
approximately correspond to the orbit of Neptune.  Therefore, estimating the occurrence rate 
of multiplanetary systems with cold planets will provide an important constraint on the planet 
formation mechanism.

\begin{figure}
\includegraphics[width=\columnwidth]{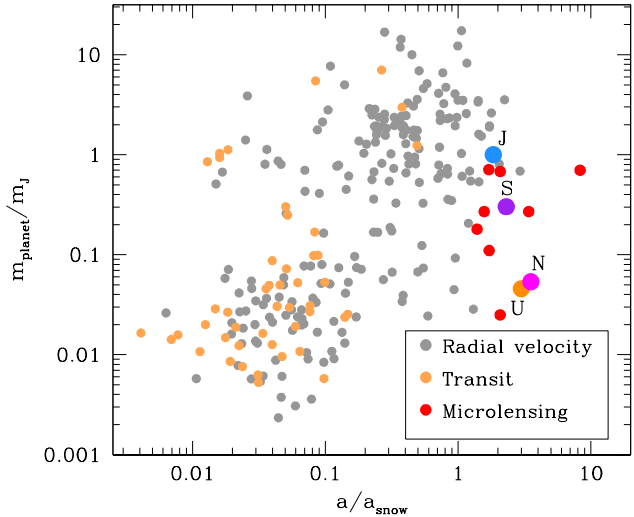}
\caption{
Distribution of planets in multiple planetary systems on the $a/a_{\rm snow}$--$m_{\rm planet}$ 
plane.  The plot is based on 257 planets in 109 systems with known planet masses.  Here 
$a_{\rm snow}$ represents the snow line.  In order to compare the planets with those of the 
Solar system, we present the locations of Jupiter (``J''), Saturn (``S''), Uranus (``U''), 
and Neptune (``N'').  The colors of points indicate the detection methods that are marked 
in the inset.
\smallskip
}
\label{fig:one}
\end{figure}

Microlensing provides an important tool to detect multiplanetary systems with cold, 
wide-orbit planets.  There currently exist 1325 planets in 657 known multiplanetary 
systems by the time of writing this 
paper.\footnote{https://exoplanetarchive.ipac.caltech.edu/index.html} Most of transit 
planets and majority of RV planets in multiplanetary systems are hot or warm planets 
located within the snow lines of the systems, and many of these planets are believed 
to have migrated from the place of their formation to their current locations by 
various dynamical mechanisms \citep{Lin1996, Ward1997, Murray1998}.  Microlensing, 
on the other hand, is sensitive to cold planets that are likely to have formed in 
situ and have not undergone large-scale migration, and thus construction of an unbiased 
sample of planets in this region is important for the investigation of giant planet 
formation.  The high sensitivity of the microlensing method to wide-orbit planets in 
multiple planetary systems is shown in the distribution of planets on the 
$a/a_{\rm snow}$--$m_{\rm planet}$ plane presented in Figure~\ref{fig:one}.  Here $a$ 
and $m_{\rm planet}$ represent the semi-major axis and mass of the planet, respectively.  
Furthermore, the microlensing method does not rely on the luminosity of the host star.  
This makes the microlensing method a useful tool for investigating multiplanetary systems 
with faint host stars, for which the sensitivity of the other methods is low.  This is 
particularly important for probing exoplanet populations, given that M dwarfs are the 
most common stars in the Galaxy.

\begin{figure}
\includegraphics[width=\columnwidth]{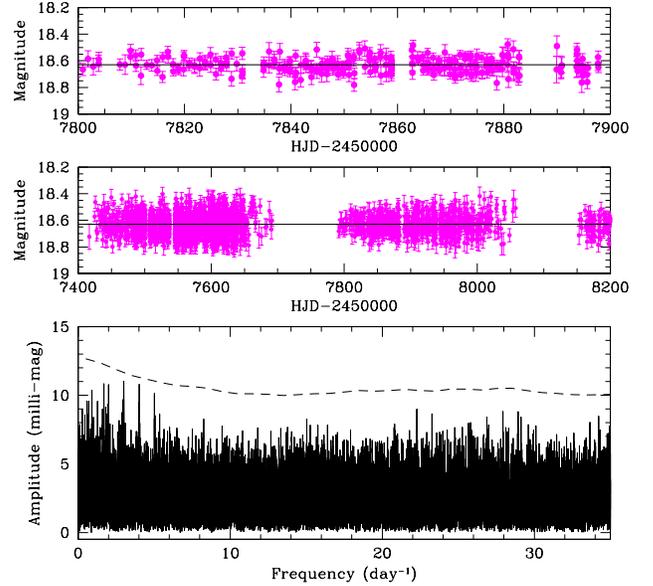}
\caption{
Baseline light curve.
The upper and lower panels show 100 day and $\sim 2.2$ year baseline observed by 
the OGLE survey, respectively.
\smallskip
}
\label{fig:two}
\end{figure}

There exist four reported multiple-planet systems detected using the microlensing 
method.  The first system, OGLE-2006-BLG-109L, is composed of a host with approximately 
half of the solar mass and two planets with masses of $\sim 0.71~M_{\rm J}$ and 
$\sim 0.27~M_{\rm J}$ and orbital separations of $\sim 2.3$~au and $\sim 4.6$~au, 
and thus the system resembles a scaled version of our solar system in that the 
mass ratio, separation ratio, and the equilibrium temperatures of the planets are 
similar to those of Jupiter and Saturn \citep{Gaudi2008, Bennett2010}.  For 
OGLE-2012-BLG-0026L, two planets with masses of $\sim 0.145~M_{\rm J}$ and $\sim 0.86~M_{\rm J}$  
are orbiting a host star with a mass $\sim 1.06~M_\odot$ \citep{Han2013, Beaulieu2016}. 
For this system, there exist 4 degenerate solutions in the interpretation of the projected 
planet-host separations, but the separations of the individual planets are beyond the snow 
line in all solutions, being $\sim 4.0$~au and $\sim 4.8$~au for the best-fit solution.  
From the statistical arguments and dynamical analysis of the orbital configuration, 
\citet{Madsen2019} argued 
that the two massive planets in OGLE-2012-BLG-0026L were likely in a resonance configuration.  
For OGLE-2014-BLG-1722L, two planets with masses of $\sim 0.18~M_{\rm J}$ and 
$\sim 0.27~M_{\rm J}$ are orbiting a late-type star with a mass of $\sim 0.4~M_\odot$.  
The projected separations from the host are $\sim 1.5$ au for the first planet and 
$\sim 1.7$~au or $\sim 2.7$~au for the second planet, and thus both planets are also 
located beyond the snow line \citep{Suzuki2018}.  OGLE-2018-BLG-0532L is the most 
recently reported candidate system, in which two planets with masses of $\sim 8~M_\oplus$ 
and $\sim 0.7~M_{\rm J}$ are located around an M-dwarf host having a mass of 
$\sim 0.25~M_\odot$ with projected separations from the host of $\sim 1.4$~au and 5.6~au, 
respectively \citep{Ryu2019}.  From the detection efficiency analysis of the MOA 
\citep{Suzuki2016} and $\mu$FUN \citep{Gould2010} surveys for the two microlensing 
multiplanetary systems OGLE-2006-BLG-109L and OGLE-2014-BLG-1722L, \citet{Suzuki2018} 
estimated that the occurrence rate of systems with multiple cold gas giant systems was 
$6\% \pm 2\%$.  We note that the multiple-planet signatures in the lensing light curves 
are securely detected for OGLE-2006-BLG-109L and OGLE-2012-BLG-0026L, but the signatures 
for OGLE-2014-BLG-1722L and OGLE-2018-BLG-0532L are less reliable.

In this paper, we report a new multiple-planet system discovered by analyzing the 
combined data of the microlensing event OGLE-2018-BLG-1011 obtained by five lensing 
surveys.  We describe data acquisition and processing in Section~\ref{sec:two}.  
In Sections~\ref{sec:three} through \ref{sec:five}, we describe the detailed 
procedure of analysis leading to the interpretation of the lens as a multiple-planet 
system.  In Sections~\ref{sec:six} and \ref{sec:seven}, we characterize the source 
of the event and estimate the physical parameters of the lens system, respectively.  
We summarize the results and conclude in Section~\ref{sec:eight}.

\begin{figure}
\includegraphics[width=\columnwidth]{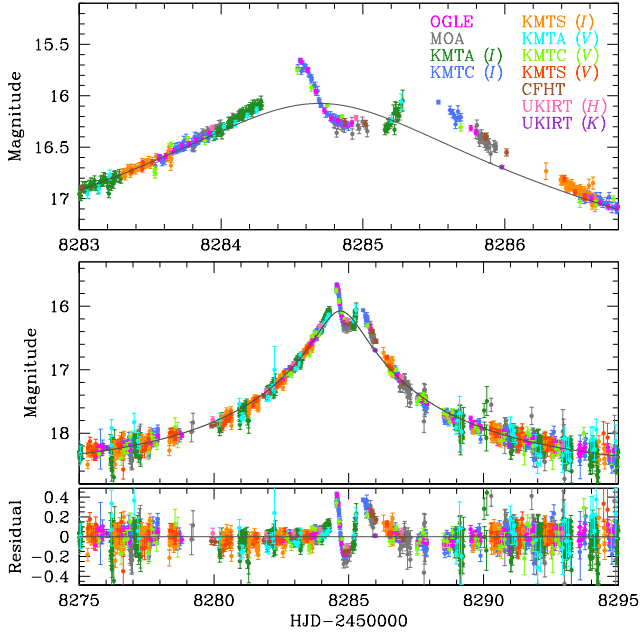}
\caption{
Light curve of the microlensing event OGLE-2018-BLG-1011. The upper panel shows 
the enlarged view of the anomaly region around the peak. The curve superposed 
on the data points is the model curve based on the point-source point-lens 
interpretation.
\smallskip
}
\label{fig:three}
\end{figure}

\section{Observation and Data}\label{sec:two}

The source star of the microlensing event OGLE-2018-BLG-1011 is located toward 
the Galactic bulge with equatorial coordinates 
$({\rm R.A.}, {\rm decl.})_{\rm J2000}=(17:56:03.36, -29:04:58.4)$, which correspond 
to Galactic coordinates of $(l, b)=(1.04^\circ, -2.04^\circ)$. 
In the middle panel of Figure~\ref{fig:two}, we present the $\sim 2.2$ year baseline 
light curve observed by the OGLE survey.  The top panel, showing the light curve 
for 100 days during $7800\leq {\rm HJD}^\prime\equiv {\rm HJD}-2450000\leq 7900$,
is presented to check the short-term variability of the source brightness.
It is found that the baseline magnitude, $I_{\rm base}\sim 18.63$, is stable
and the light curve does not show noticeable variability.
We further investigate the source variability by computing the power spectrum
using the PERIOD04 code of \citet{Lenz2005}.
The power spectrum is presented in the bottom panel. 
The dashed line represents the limit with the signal-to-noise ratio ${\rm S/N}= 4$,
which is the empirically proposed threshold for variability \citep{Breger1993}.
The spectrum shows no periodic variability greater than the imposed threshold, 
indicating that the source flux has been stable.

The brightening of the source star induced by lensing was found in the early 
rising stage of the event by the Optical Gravitational Lensing Experiment 
\citep[OGLE:][]{Udalski2015b}, and the discovery was notified to the microlensing 
community on 2018-06-07 (${\rm HJD}^\prime\sim 8277$).  
The event was in the OGLE-IV BLG505.23 field that was monitored with a cadence 
of 1 day using the 1.3 m telescope located at the Las Campanas Observatory in 
Chile. Most OGLE data were acquired in $I$ band, and some $V$-band images were 
obtained for the source color measurement.

The event was also located in the fields of two other major lensing surveys, namely 
the Microlensing Observations in Astrophysics \citep[MOA:][]{Bond2001, Sumi2003} 
and Korea Microlensing Telescope Network \citep[KMTNet:][]{Kim2016}.
The MOA observations of the event were conducted in a customized broad $R$ band using 
the 1.8~m telescope located at the Mt.~John Observatory in New Zealand.  The cadence 
of the MOA survey was $\sim 10$/night in a survey mode and reached up to 40/night 
when the anomaly in the lensing light curve was in progress.  The event was designated 
as MOA-2018-BLG-182 in the list of the ``MOA Transient Alerts''
page.\footnote{http://www.massey.ac.nz/$\sim$iabond/moa/alert2018/alert.php.}

\begin{figure}
\includegraphics[width=\columnwidth]{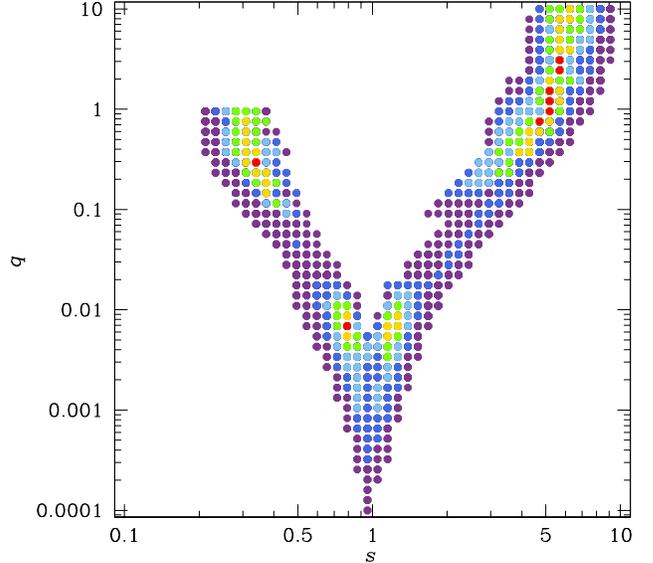}
\caption{
$\Delta\chi^2$ distribution on the plane of the binary-lensing parameters $s$ and $q$. 
The color coding indicates points within $1n\sigma$ (red), $2n\sigma$ (yellow), $3n\sigma$ 
(green), $4n\sigma$ (cyan), $5n\sigma$ (blue), and $6n\sigma$ (purple) with respect to 
the best-fit value and $n=10$.
\smallskip
}
\label{fig:four}
\end{figure}

The KMTNet survey utilizes three identical 1.6~m telescopes that are globally 
distributed for continuous coverage of lensing events.  The individual telescopes 
are located at the Siding Spring Observatory in Australia (KMTA), Cerro Tololo 
Interamerican Observatory in Chile (KMTC), and the South African Astronomical 
Observatory in South Africa (KMTS).  The event was in the two overlapping fields 
(BLG02 and BLG042) of the KMTNet survey and images were obtained mainly in $I$ 
band with occasional $V$-band data acquisition.  We use $V$-band data not only 
for the source color measurement but also for the light curve analysis to maximize 
the coverage of the short-term anomaly that appeared in the peak of the event 
light curve.  The observational cadence of the KMTNet survey varied depending 
on the observation site, ranging 4/hr for the KMTC telescope and 6/hr for the 
KMTA and KMTS telescopes.  The event was detected by the KMTNet ``event finder'' 
\citep{Kim2018} and was named as KMT-2018-BLG-2122.

Besides the three major microlensing surveys, the event was also observed by 
two lower-cadence surveys conducted by utilizing  the Canada-France-Hawaii 
Telescope (CFHT) \citep{Zang2018} and the 3.8~m United Kingdom Infrared Telescope 
(UKIRT) \citep{Shvartzvald2017}.  The CFHT data include 62 points in the time 
range of $8256 \lesssim {\rm HJD}^\prime \lesssim 8346$.  The UKIRT observations 
were conducted with a daily cadence in $H$ band and occasional observations in 
$K$ band.  The CFHT data include 62 points obtained during 
$8256 \lesssim {\rm HJD}^\prime \lesssim 8346$ and the UKIRT data include 49 
$H$-band and 10 $K$-band data points acquired during 
$7864 \lesssim {\rm HJD}^\prime \lesssim 8336$ and 
$7882 \lesssim {\rm HJD}^\prime \lesssim 8329$, respectively.

Data sets obtained by the OGLE and MOA surveys were released almost in real time 
through the ``Early Warning System''\footnote{http://ogle.astrouw.edu.pl/ogle4/ews/ews.html.} 
and ``MOA Transient Alerts'' pages.  This in turn facilitated monitoring of the real-time 
evolution of the event.  On 2018-06-16 (${\rm HJD}^\prime\sim 8285.5$), D.~Suzuki of the 
MOA group noticed a deviation of the light curve from a point-source point-lens model and 
issued an anomaly alert.  With this alert, the MOA group increased their observation cadence. 
See the light curve of the event and the anomaly around the peak presented in Figure~\ref{fig:three}. 
The alert triggered real-time analysis of the light curve and a series of models, all of which 
were based on binary-lens interpretations, were released by V.~Bozza, A.~Cassan, D.~Bennett, 
and Y.~Hirao during the progress of the anomaly.  Although results were not circulated, the 
first author of this paper (C.~Han) was also conducting analysis of the event with the progress 
of the event.  When the anomaly ended, however, it was found that none of these models could 
fully explain the observed anomaly.  The fact that the real-time analyses done by five people 
using separate, independently-written softwares reached the same conclusion that a binary-lens 
interpretation cannot properly explain the observed data strongly suggests that one needs an 
interpretation of the event that is different from the binary-lens explanation.

We note that there are gaps in the data during the anomaly despite the coverage by five surveys.  
The gaps centered at ${\rm HJD}^\prime \sim 8284.4$ and 8285.4 could have been covered by KMTS 
telescope located in Africa and the gap centered at ${\rm HJD}^\prime \sim 8286.2$ could have 
been observed by KMTA telescope located in Australia.  Unfortunately, no data could be obtained 
because of the poor weather conditions of the sites.

\begin{deluxetable}{lll}
\tablecaption{Photometric Uncertainty Rescaling Parameters\label{table:one}}
\tablewidth{240pt}
\tablehead{
\multicolumn{1}{c}{Data set}      &
\multicolumn{1}{c}{$k$}      & 
\multicolumn{1}{c}{$\sigma_0$ (mag)}  
}
\startdata                 
OGLE                  &  1.170      &   0.020      \\
MOA                   &  1.490      &   0.020      \\
KMTA ($I$, BLG02)     &  1.523      &   0.020      \\
KMTA ($I$, BLG42)     &  2.170      &   0.020      \\
KMTC ($I$, BLG02)     &  1.086      &   0.030      \\
KMTC ($I$, BLG42)     &  1.584      &   0.010      \\
KMTS ($I$, BLG02)     &  1.396      &   0.020      \\
KMTS ($I$, BLG42)     &  1.261      &   0.030      \\
KMTA ($V$, BLG02)     &  2.017      &   0.010      \\
KMTA ($V$, BLG42)     &  2.135      &   0.010      \\
KMTC ($V$, BLG02)     &  1.470      &   0.025      \\
KMTC ($V$, BLG42)     &  1.421      &   0.005      \\
KMTS ($V$, BLG02)     &  1.288      &   0.010      \\
KMTS ($V$, BLG42)     &  1.253      &   0.010      \\
CFHT                  &  0.806      &   0.040      \\
UKIRT ($H$)           &  0.500      &   0.030      \\
UKIRT ($K$)           &  0.224      &   0.030      
\enddata                            
\tablecomments{
The notations in the parentheses of the KMTNet data sets 
represent the passbands and fields of observation.
\smallskip
}
\end{deluxetable}

\begin{deluxetable}{lllc}
\tablecaption{Comparison of models\label{table:two}}
\tablewidth{240pt}
\tablehead{
\multicolumn{3}{c}{Solution}       &
\multicolumn{1}{c}{$\chi^2$}
}
\startdata
2L1S  & binary               & $s<1.0$                 & 8545.6   \\
      &                      & $s>1.0$                 & 8470.6   \\
      & planetary            & $s<1.0$                 & 8439.7   \\ 
      &                      & $s>1.0$                 & 8477.0   \\
\hline
1L2S  &                      &                         & 10853.6  \\
\hline 
2L2S  &                      &                         & 8047.3   \\
\hline
3L1S  & Planet-binary        &  $s_2<1.0$, $s_3<1.0$   & 7825.4   \\
      &                      &  $s_2<1.0$, $s_3>1.0$   & 7825.5   \\
      &                      &  $s_2>1.0$, $s_3<1.0$   & 7865.2   \\
      &                      &  $s_2>1.0$, $s_3>1.0$   & 7882.9   \\
\hline                                                   
      & Multiple-planet~(I)  &  $s_2<1.0$              & 7783.8   \\
      &                      &  $s_2>1.0$              & 7790.7   \\
\hline
      & Multiple-planet~(II) &  $s_2<1.0$, $s_3<1.0$   & 7718.0   \\
      &                      &  $s_2<1.0$, $s_3>1.0$   & 7761.6   \\
      &                      &  $s_2>1.0$, $s_3<1.0$   & 7717.7   \\
      &                      &  $s_2>1.0$, $s_3>1.0$   & 7756.5   

\enddata
\tablecomments{For the 3L1S solutions, $s_2$ and $s_3$ represent the 
normalized separation between $M_1$--$M_2$ and $M_1$--$M_3$, respectively, 
and $M_1$ represents the primary lens and $M_2$ and $M_3$ denote the companions.
\bigskip
}
\end{deluxetable}

In our analysis of the event, we use photometry data processed with the codes developed 
by the individual survey groups: \citet{Wozniak2000} (OGLE), \citet{Bond2001} (MOA) and 
\citet{Albrow2017} (KMTNet).  All of these codes are based on the difference imaging 
technique of \citet{Alard1998}.  For a subset of KMTNet data, additional photometry is 
processed using the pyDIA photometry \citep{Albrow2017} to measure the source color.  
For the use of multiple data sets obtained by different groups and processed by using 
different photometry codes, we normalize the photometric measurement uncertainties of 
the data sets following the procedure of \citet{Yee2012}, in which the photometric 
uncertainties are rescaled by 
\begin{equation}
\sigma=k(\sigma_0^2+\sigma_{\rm min}^2)^{1/2}.
\label{eq1}
\end{equation}
The quadratic term $\sigma_0$ is added so that the cumulative distribution of $\chi^2$ 
ordered by magnification is approximately linear.  This process ensures the dispersion 
of data points consistent with error bars.  The coefficient ``$k$'' is a factor used for  
rescaling the errors so that $\chi^2$ per degree of freedom ($\chi^2/{\rm dof}$) for each 
data set becomes unity.  The latter process is needed to prevent each data set from being 
under or over-represented compared to other data sets.  In Table~\ref{table:one}, we present 
the values of $k$ and $\sigma_0$.

\begin{deluxetable*}{l|ll|ll|l|l}
\tablecaption{Lensing parameters of 2L1S, 1L2S, and 2L2S models\label{table:three}}
\tablewidth{500pt}
\tablehead{
\multicolumn{1}{c|}{Parameter}   &
\multicolumn{2}{c|}{2L1S (Binary)}        &
\multicolumn{2}{c|}{2L1S (Planet)}        &
\multicolumn{1}{c|}{1L2S}        & 
\multicolumn{1}{c}{2L2S}        \\
\multicolumn{1}{c|}{}            &
\multicolumn{1}{c}{($s<1$)}     &
\multicolumn{1}{c|}{($s>1$)}     &
\multicolumn{1}{c}{($s<1$)}     &
\multicolumn{1}{c|}{($s>1$)}     &
\multicolumn{1}{c|}{}            & 
\multicolumn{1}{c}{}            
}
\startdata                 
$\chi^2$                           &  8546.6                &  8470.6                &  8439.7                             &   8477.0                            &  10853.6                     &  8047.3                                    \\  
$t_{0,1}$ (${\rm HJD}^\prime$)     &  $8284.605 \pm 0.003$  &  $8284.594 \pm 0.002$  &  $8284.751 \pm 0.003$               &   $8284.755 \pm 0.003$              &  $8284.439  \pm    0.004$    &  $8284.646 \pm 0.004 $                     \\
$u_{0,1}$                          &  $   0.032 \pm 0.001$  &  $   0.022 \pm 0.001$  &  $   0.062 \pm 0.001$               &   $   0.063 \pm 0.001$              &  $   0.061  \pm    0.001$    &  $   0.063 \pm 0.001 $                     \\
$t_{0,2}$ (${\rm HJD}^\prime$)     &  --                    &   --                   &   --                                &    --                               &  $8285.686  \pm    0.011$    &  $8285.403 \pm 0.007 $                     \\
$u_{0,2}$                          &  --                    &   --                   &   --                                &    --                               &  $   0.027  \pm    0.001$    &  $   0.011 \pm 0.001 $                     \\
$t_{\rm E}$ (days)                 &  $12.73 \pm 0.15$      &  $17.53 \pm  0.25$     &  $10.61 \pm  0.11$                  &   $10.58 \pm  0.11$                 &  $  12.69   \pm    0.16 $    &  $  11.05  \pm 0.08  $                     \\
$s$                                &  $0.338 \pm 0.003$     &  $5.097 \pm  0.070$    &  $0.807 \pm  0.004$                 &   $1.169 \pm  0.006$                &  --                          &  $   0.784 \pm 0.002 $                     \\
$q$                                &  $0.301 \pm 0.015$     &  $1.108 \pm  0.096$    &  $(6.98 \pm 0.15)\times 10^{-3}$    &   $(7.03 \pm 0.17)\times 10^{-3}$   &  --                          &  $   (6.61 \pm 0.09)\times 10^{-3}$        \\
$\alpha$ (rad)                     &  $0.246 \pm 0.005$     &  $0.235 \pm  0.005$    &  $4.434 \pm  0.005$                 &   $4.440 \pm  0.005$                &  --                          &  $   4.253 \pm 0.008 $                     \\
$\rho_1$ ($10^{-3}$)               &  --                    &   --                   &   --                                &    --                               &  --                          &  --                                        \\
$\rho_2$ ($10^{-3}$)               &  --                    &   --                   &   --                                &    --                               &  $   7.58   \pm    3.07 $    &  $   0.80  \pm 0.13  $                     \\
$q_{F,I}$                          &  --                    &   --                   &   --                                &    --                               &  $   0.24   \pm    0.01 $    &  $   0.071 \pm 0.003 $                     
\enddata  
\tablecomments{${\rm HJD}^\prime={\rm HJD-2450000}$.
\bigskip
}
\end{deluxetable*}

\section{Binary-lens (2L1S) interpretations}\label{sec:three}

The light curve of the event exhibits an anomaly that is characterized by two bumps 
near the peak.  We start the analysis of the anomaly with the ``2L1S'' interpretation, 
in which the event is produced by a lens composed of two masses, ``2L,'' deflecting  
the light from a single source, ``1S.''

It is known that a double-bump feature in the peak region of a lensing light curve 
can be produced by two major channels of 2L1S events \citep{Han2008}.  One channel 
is the case, in which the source trajectory approaches the two neighboring 
cusps of the central caustic produced by a binary lens composed of similar masses with a 
projected separation between the lens components significantly smaller (close binary) 
or wider (wide binary) than the Einstein radius $\thetae$ of the lens system.  The 
other channel is the case, in which the source approaches the back-end cusp 
of the central caustic produced by a planetary lens system with a mass ratio between 
the lens components $q\ll 1$ and a star-planet separation similar to $\thetae$ 
\citep{Choi2012, Park2014, Bozza2016}.  We refer to the former and latter channels 
as the ``binary'' and ``planetary'' channels, respectively.

\begin{figure}
\includegraphics[width=\columnwidth]{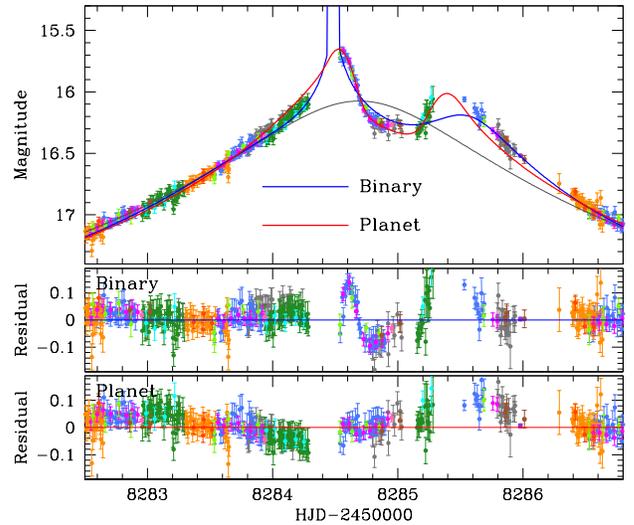}
\caption{
Model light curves of the two solutions obtained for the 2L1S interpretation: 
binary (blue curve: $s < 1.0$ (close solution); cf., 
Table~\ref{table:three}) and planetary (red curve: $s < 1.0$ close 
solution); cf., Table~\ref{table:three}) solutions.  The two lower panels 
show the residuals from the individual models.  
The lens-system configurations corresponding to the individual solutions are 
presented in Fig.~\ref{fig:six}.
\bigskip
}
\label{fig:five}
\end{figure}

For the 2L1S analysis, we first conduct a dense grid search for the lensing parameters 
$s$ and $q$, which represent the binary separation normalized to $\thetae$ and the mass 
ratio between the lens components, respectively. 
Considering that central perturbations can be produced either by a planetary companion 
with a low mass ratio or a binary companion with very wide or close separations from 
the primary, we set the ranges of $s$ and $q$ for the grid search wide enough to check 
both the planetary and binary solutions.  
The initial grid search is done in the ranges of $-1.0\leq \log s \leq 1.0$ and 
$-4.0\leq \log q \leq 1.0$ for the separation and mass ratio, respectively, with 
50 grids for each range. 
We then identify local minima in the distribution of $\Delta\chi^2$ on 
the $s$--$q$ plane. For the individual local solutions, we 
gradually narrow down the ranges of the parameter space in the grid search. 
Besides these lensing parameters, a basic 
2L1S modeling requires additional parameters, including the time of the closest 
lens-source approach, $t_0$, the lens-source separation at that time, $u_0$ (normalized 
to $\thetae$), the Einstein timescale, $t_{\rm E}$, and the source trajectory angle 
with respect to the binary-lens axis, $\alpha$. In the case when the source crosses 
the caustic formed by the binary lens, the lensing light curve is affected by 
finite-source effects and one needs an additional parameter of $\rho$, which represents 
the source radius $\theta_*$ normalized to $\thetae$, i.e., $\rho=\theta_*/\thetae$, 
to account for these effects.  (MCMC) method.  
In computing finite-source magnifications, we consider surface-brightness variation 
of the source stars caused by the limb darkening.  The stellar type of the source 
turns out to be a G-type turn-off star (Section~\ref{sec:six}) and thus we adopt a 
linear limb-darkening coefficient of $\Gamma_I=0.5$.  For a given set of 
$(s, q)$, we search for the other lensing parameters using a downhill approach based 
on the Markov Chain Monte Carlo (MCMC) method.

In Figure~\ref{fig:four}, we present the distribution of $\Delta\chi^2$ from the 
best-fit model on the $s$--$q$ parameter plane obtained from the initial grid search. 
We find that there exist four local solutions.  For one pair of the local solutions, 
the mass ratios are $q\sim 1$, and thus the solutions correspond to binary-lens 
solutions.  For the other pair, the mass ratios are $q < 10^{-2}$, indicating that 
they correspond to planetary mass lens solutions.  For each pair, we find that 
there exist two solutions with $s<1$ (close solution) and $s>1$ 
(wide solution), which are generated  by the close/wide degeneracy 
\citep{Griest1998, Dominik1999}.

\begin{figure}
\includegraphics[width=\columnwidth]{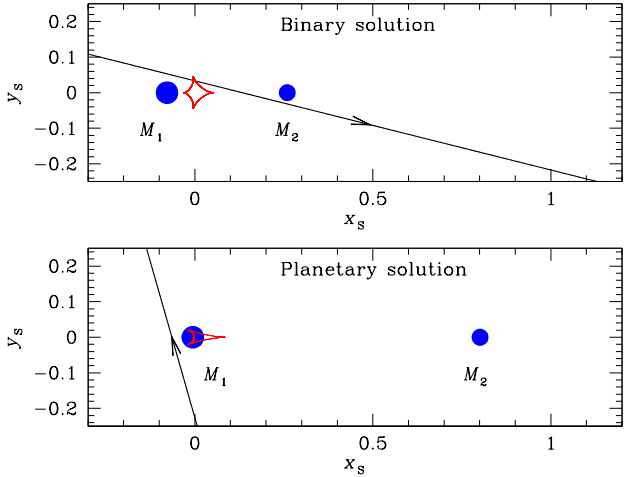}
\caption{
Lens-system configurations of the 2L1S solutions.  The upper and lower panels correspond 
to the ``binary'' (with $s<1.0$) and ``planetary'' (with $s<1.0$) solutions, respecively.  
In each panel, the line with an arrow represents the source trajectory and the red close 
curve is the caustic.  The blue dots marked by $M_1$ and $M_2$ denote the positions of 
the binary lens components. 
\bigskip
}
\label{fig:six}
\end{figure}

In Figure~\ref{fig:five}, we present the model light curves corresponding to the 
``binary'' (blue curve) and ``planetary'' (red curve) 2L1S solutions superposed on 
the observed data, together with the residuals from the models (presented in the 
lower two panels).  We note that the individual solutions are further refined based 
on the local minima obtained from the grid search by allowing all lensing parameters, 
including $s$ and $q$,  to vary.  In Table~\ref{table:two}, we compare the $\chi^2$ 
values of the four 2L1S solutions.  In Table~\ref{table:three}, we list the best-fit 
lensing parameters of the individual 2L1S solutions.  We also note that the model 
light curves corresponding to the close and wide solutions of each of the binary and 
planetary solution pairs are similar to each other and thus in Figure~\ref{fig:five} 
we present the one yielding a better fit to the data.  
In Figure~\ref{fig:six}, we present the lens-system configurations, which 
represent the source trajectory with respect to the caustic, of the 
binary (upper panel) and planetary (lower panel) solutions. 
We note that the presented configurations are for the close ($s<1.0$) solutions.
From the fits of the individual models to the observed data, it is found that the binary-lens 
solution leaves substantial residuals in the region 
$8284.5 \lesssim {\rm HJD}^\prime \lesssim 8285.6$, and the planetary solution leaves 
residuals in the region $8285.6 \lesssim {\rm HJD}^\prime \lesssim 8286.0$.  
These residuals indicate that neither the binary nor the planetary 2L1S solutions adequately 
describe the observed data and suggest that a new interpretation of the light curve is 
required.

We additionally check whether the fit further improves by considering the microlens-parallax 
\citep{Gould1992} and/or the lens-orbital \citep{Dominik1998, Ioka1999} effects.  It is found 
that the improvement by these higher-order effects is negligible, mainly due to the short 
timescale, $t_{\rm E}\sim 12$~days, of the event.

\begin{deluxetable*}{lllll}
\tablecaption{Lensing parameters of ``planet-binary'' 3L1S solutions\label{table:four}}
\tablewidth{500pt}
\tablehead{
\multicolumn{1}{c}{Parameter}                 &
\multicolumn{1}{c}{$s_2<1.0$, $s_3<1.0$}      & 
\multicolumn{1}{c}{$s_2<1.0$, $s_3>1.0$}      & 
\multicolumn{1}{c}{$s_2>1.0$, $s_3<1.0$}      & 
\multicolumn{1}{c}{$s_2>1.0$, $s_3>1.0$}  
}
\startdata                 
$\chi^2$                   &  7825.4                              &  7825.5                              &  7865.2                             &  7882.9                                 \\
$t_0$ (${\rm HJD}^\prime$) &  8284.557 $\pm$ 0.006                &  8284.564 $\pm$ 0.005                &  8284.557 $\pm$ 0.004               &  8284.549 $\pm$ 0.003                   \\
$u_0$                      &  0.033 $\pm$ 0.001                   &  0.034 $\pm$ 0.001                   &  0.023 $\pm$ 0.001                  &  0.023 $\pm$ 0.001                      \\
$t_{\rm E}$ (days)         &  12.80 $\pm$ 0.14                    &  12.70 $\pm$ 0.15                    &  17.12 $\pm$ 0.20                   &  16.59 $\pm$ 0.12                       \\
$s_2$                      &  0.357 $\pm$ 0.004                   &  0.354 $\pm$ 0.004                   &  4.872 $\pm$ 0.073                  &  4.559 $\pm$ 0.039                      \\
$q_2$                      &  0.263 $\pm$ 0.012                   &  0.273 $\pm$ 0.012                   &  0.977 $\pm$ 0.079                  &  0.703 $\pm$ 0.035                      \\
$\alpha$ (rad)             &  0.282 $\pm$ 0.0054                  &  0.283 $\pm$ 0.005                   &  0.268 $\pm$ 0.004                  &  0.265 $\pm$ 0.004                      \\
$s_3$                      &  0.812 $\pm$ 0.015                   &  1.262 $\pm$ 0.023                   &  0.896 $\pm$ 0.025                  &  1.002 $\pm$ 0.006                      \\
$q_3$                      &  $(2.81 \pm 0.45) \times 10^{-3}$    &  $(2.28 \pm  0.46) \times 10^{-3}$   &  $(1.71 \pm 0.26) \times 10^{-3}$   &  $(2.28 \pm  0.26) \times 10^{-3}$      \\
$\psi$ (rad)               &  1.259 $\pm$ 0.006                   &  1.299 $\pm$ 0.005                   &  1.371 $\pm$ 0.005                  &  1.361 $\pm$ 0.004                      
\enddata                            
\tablecomments{${\rm HJD}^\prime={\rm HJD-2450000}$. 
\bigskip
}
\end{deluxetable*}

\section{Binary-source (1L2S and 2L2S) Interpretation}\label{sec:four}

\citet{Gaudi1998} pointed out the possibility that a short-term perturbation, 
which was the main feature of a planetary microlensing signal produced by major-image 
perturbations, could be reproduced by a subset of binary-source events, in which a 
single-mass lens, 1L, passed close to the fainter member of the binary source, 2S.  
The degeneracy between 2L1S and 1L2S interpretations can often be severe, as 
demonstrated in the cases OGLE-2002-BLG-055 \citep{Gaudi2004}, MOA-2012-BLG-486 
\citep{Hwang2013}, and OGLE-2014-BLG-1186 \citep{Dominik2019}.  Besides the 
short-term anomaly considered by \citet{Gaudi1998}, it was shown that the 
2L1S/1L2S degeneracy could extend to various cases of planetary lens system 
configurations from the analysis of anomalies in the cases of OGLE-2016-BLG-0733 
by \cite{Jung2017} and KMT-2017-BLG-0962 and KMT-2017-BLG-1119 by \citet{Shin2019}.

\begin{figure}
\includegraphics[width=\columnwidth]{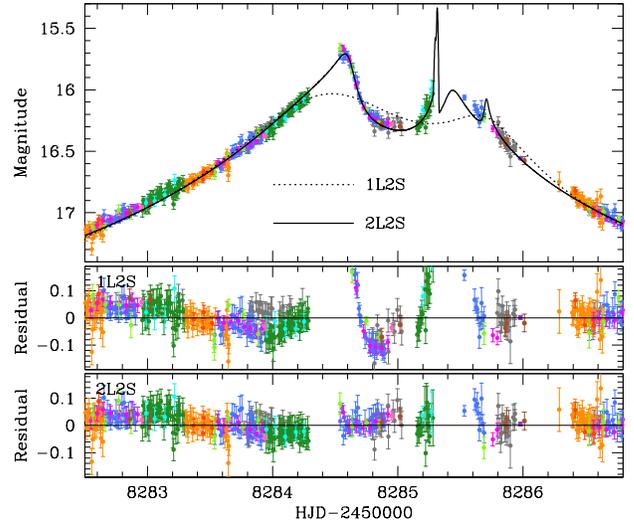}
\caption{
Model light curves obtained for the 1L2S (dotted curve) and 2L2S (solid curve) 
interpretations.  The two lower panels show the residuals of the individual models.  
The inset in the top panel shows the lens-system configuration of the 2L2S solution.  
The black lines with arrows marked by ``${\rm S}_1$'' and ``${\rm S}_2$'' represent 
the trajectories (lines with arrows) of the binary-source components with respect 
to the caustic (red closed figure).
\bigskip
}
\label{fig:seven}
\end{figure}

We check the binary-source interpretation of the anomaly in the light curve of 
OGLE-2018-BLG-1011 by conducting a 1L2S modeling of the observed light curve.  
The addition of the source companion requires the inclusion of additional parameters 
in modeling.  These parameters are $t_{0,2}$, $u_{0,2}$, and $q_F$, which 
represent the time of the closest lens approach to the source companion, the 
impact parameter to the companion, and the flux ratio between the source stars, 
respectively.  We note that finite-source effects are considered in the modeling.  
From this modeling, we find that the 1L2S solution provides a poorer fit than the 
best-fit 2L1S solutions, by $\Delta\chi^2= 2313.1$ and $2418.9$ compared to the 
binary and planetary 2L1S solutions, respectively, and thus we reject the solution.  
In Figure~\ref{fig:seven}, we present the best-fit 1L2S model (dotted curve 
superposed on the data points).

\begin{figure}
\includegraphics[width=\columnwidth]{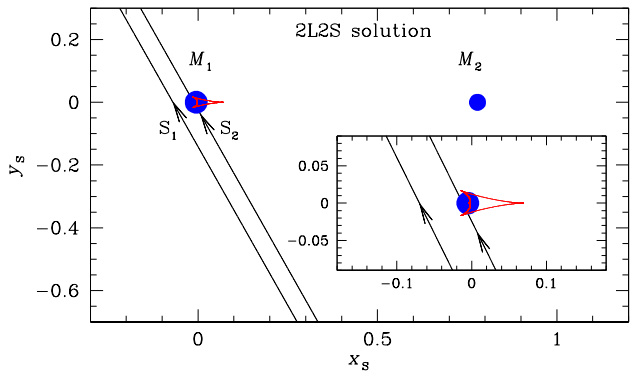}
\caption{
Lens-system configuration of the best-fit 2L2S model.
Notations are same as those in Fig.~\ref{fig:six} except that there 
are two source trajectories because the source is a binary.
The trajectories marked by $S_1$ and $S_2$ represent those of the 
brighter and fainter source stars, respectively.  
\bigskip
}
\label{fig:eight}
\end{figure}

We also check the model in which both the source and lens are binaries, 2L2S model. 
Figure~\ref{fig:seven} shows the model light curve (solid curve) of the best-fit 2L2S 
solution.  In Figure~\ref{fig:eight}, we also present the lens-system configuration 
corresponding to the model.  We note that there are two source trajectories because 
the source is a binary in the 2L2S model.  The solution provides a 
better fit than the 2L1S and 1L2S solutions, but it leaves noticeable residuals 
in the region $8282.5 \lesssim {\rm HJD}^\prime\lesssim 8283.2$.  As we will show 
in the following section, the fit of the 2L2S solution is worse than the best-fit 
solution based on other interpretation by $\Delta\chi^2=329.6$.  In 
Table~\ref{table:two}, we list the $\chi^2$ values of the 1L2S and 2L2S solutions.  
In Table~\ref{table:three}, we also list the best-fit lensing parameters of the 1L2S 
and 2L2S solutions.

\section{Triple-lens (3L1S) Interpretation}\label{sec:five}

Knowing the inadequacy of the 2L1S, 1L2S, and 2L2S solutions in describing the observed 
light curve, we then model the light curve assuming a triple-lens interpretation, in 
which the lens contains 3 components, 3L. We try 3L1S modeling because the solutions 
obtained from the 2L1S modeling partially describe the observed central perturbation 
and the residual from the 2L1S model may be explained by introducing an additional 
lens component.

Central perturbations of lensing light curves in 3L1S cases can be produced through two 
major channels. One channel is through multiple-planet systems, in which the individual 
planets located in the lensing zone can affect the magnification pattern of the central 
perturbation region \citep{Gaudi1998b}.  Among the four known multiplanetary systems 
detected by microlensing, three systems, OGLE-2006-BLG-109Lb,c, OGLE-2012-BLG-0026Lb,c, 
and OGLE-2018-BLG-0532Lab, were detected through the central perturbations induced by 
two planets. The other channel is through planet+binary systems. Similar to the central 
caustic produced by a planet, a very close or a very wide binary companion can also 
induce a small caustic in the central magnification region and thus can affect the 
magnification pattern.  Among the four known microlensing planetary systems in 
binaries\footnote{OGLE-2007-BLG-349L(AB)c \citep{Bennett2016}, OGLE-2016-BLG-0613LABb 
\citep{Han2017}, OGLE-2008-BLG-092LABb \citep{Poleski2014}, and OGLE-2013-BLG-0341LAbB 
\citep{Gould2014b}}, the circumbinary planetary system OGLE-2007-BLG-349L(AB)c was 
detected through this channel.

The lensing behavior of triple-lens systems is qualitatively different from that of 
binary-lens systems, resulting in a complex caustic structure, such as nested caustic 
and self-intersections.  The range of the critical-curve topology and the caustic 
structure of the triple lens has not yet been fully explored, making it difficult to 
analyze triple-lens events \citep{Danek2015, Danek2019}.  As a result, there are some 
events suspected to be triple-lens events, but plausible models have yet not been 
proposed, e.g., OGLE-2008-BLG-270, OGLE-2012-BLG-0442/MOA-2012-BLG-245, 
OGLE-2012-BLG-0207/MOA-2012-BLG-105, and OGLE-2018-BLG-0043/MOA-2018-BLG-033.  
In some cases, interpretations of triple-lens events can be confused with those of 
binary-lens events, as in the cases of MACHO-97-BLG-41 
\citep{Bennett1999, Albrow2000, Jung2013} and OGLE-2013-BLG-0723 
\citep{Udalski2015a, Han2016}.

Despite the diversity and complexity of the lensing behavior, triple-lens events can 
be readily analyzed for some specific lens cases.  One such a case occurs when the 
lens companions produce small perturbations with minor interference in the magnification 
pattern between the perturbations produced by the individual companions.  In this case, 
the resulting anomaly can be approximated by the superposition of the two binary anomalies, 
in which the individual primary-companion pairs act as independent 2-body systems 
\citep{Han2005}.  This binary-superposition approximation can be applied for two cases 
of triple-lens systems, in which the lens is composed of multiple planets, as demonstrated 
in the cases of OGLE-2006-BLG-109 \citep{Gaudi2008} OGLE-2012-BLG-0026 \citep{Han2013}, 
and planets in close or wide binary systems, as demonstrated in the case of OGLE-2016-BLG-0613 
\citep{Han2017}.  The 2L1S modeling of OGLE-2018-BLG-1011 yields two local solutions, with 
a planetary companion and a close/wide binary companion, respectively.  This suggests that 
the event may be the case for which the analysis based on the superposition approximation 
is valid.

\begin{figure}
\includegraphics[width=\columnwidth]{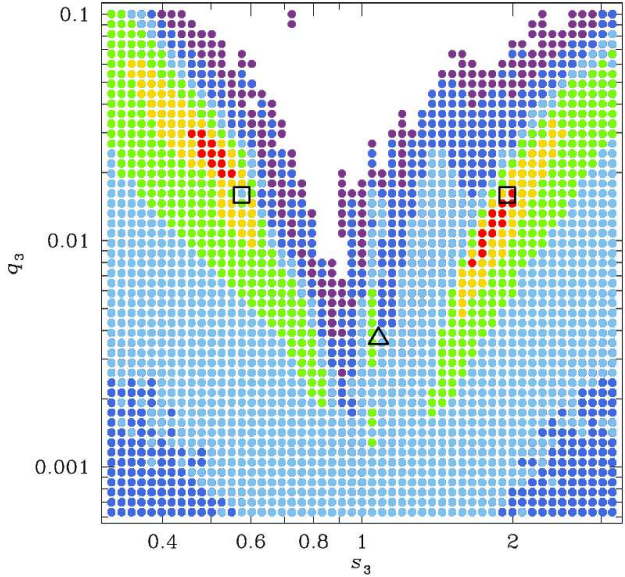}
\caption{
$\Delta\chi^2$ map in the $s_3$--$q_3$ parameter plane obtained from the grid search 
with the initial values of the ``planetary'' 2L1S solution.  The local marked by the
empty triangle dot corresponds to the ``multiple-planet (I)'' solution and the two 
locals marked by the empty square dots corresponds to the ``multiple-planet (II)'' 
solutions.  The color coding is same as that of Fig.~\ref{fig:four}.
\smallskip
}
\label{fig:nine}
\end{figure}

The 3L1S modeling is done in two steps. In the first step, we conduct a grid 
search for the separation $s_3$ and mass ratio $q_3$ between $M_1$ and $M_3$, and 
the orientation angle $\psi$ of the third body with respect to the $M_1$--$M_2$ axis. 
Here, we use the subscripts ``1''--``3'' to denote the individual lens components. 
In this search, we fix the values of ($s_2, q_2, \alpha)$ as those obtained from the 
2L1S modeling.  Because two local 2L1S solutions are found, i.e., the ``binary'' and 
``planetary'' solutions, we conduct two sets of modeling with the initial values of 
$(s_2, q_2, \alpha)$ of the ``binary'' and ``planetary'' solutions obtained from the 
2L1S modeling.  
We use the parameters of the close 2L1S solution as the initial parameters, but 
the result would not be affected with 
the use of the wide solution parameters  
because of the similarity between the model light curves of the close and wide solutions.
In Figure~\ref{fig:nine}, we present the $\Delta\chi^2$ map in the $s_3$--$q_3$ parameter 
plane obtained from the grid search with the initial values of the ``planetary'' 2L1S 
solution.
In the second step, we refine the solutions found from the grid search 
by allowing all parameters to vary.

\begin{figure*}
\epsscale{0.84}
\plotone{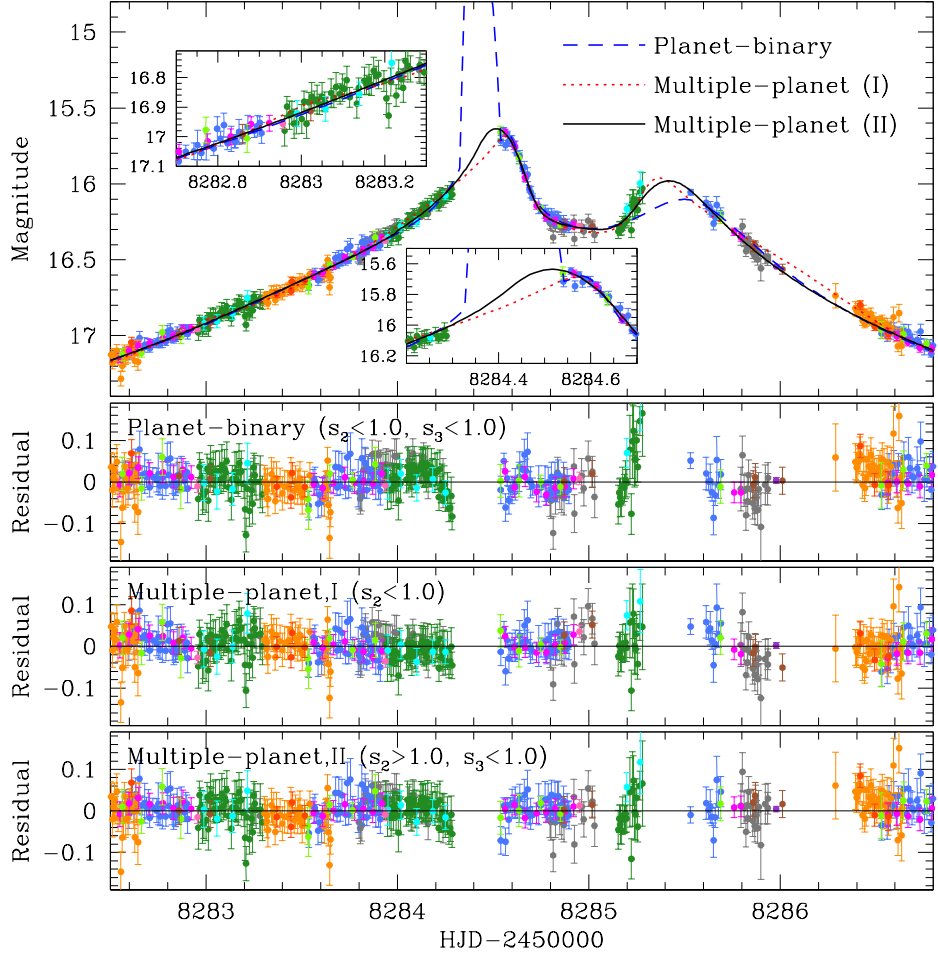}
\caption{
Model light curves of the three solutions obtained for the 3L1S interpretation:
``planet-binary'' (blue curve: $s_2 < 1, s_3 < 1$; cf., Table~\ref{table:four}), 
``multiple-planet~(I)'' (red curve: $s_2 < 1$; cf., Table~\ref{table:five}),
``multiple-planet~(II)'' (black curve: $s_2 > 1, s_3 < 1$; cf., Table~\ref{table:six}).
The three lower panels show the residuals from the individual models.
The upper left inset in the top panel shows the enlarged view of the source star's 
cusp approach at around ${\rm HJD}^\prime\sim 8283.0$ according to the ``multiple-planet~(I)'' 
solution, and the lower middle inset shows the zoom of the caustic-crossing region 
according to the ``planet-binary'' solution.  
\bigskip
}
\label{fig:ten}
\end{figure*}

\begin{figure}
\includegraphics[width=\columnwidth]{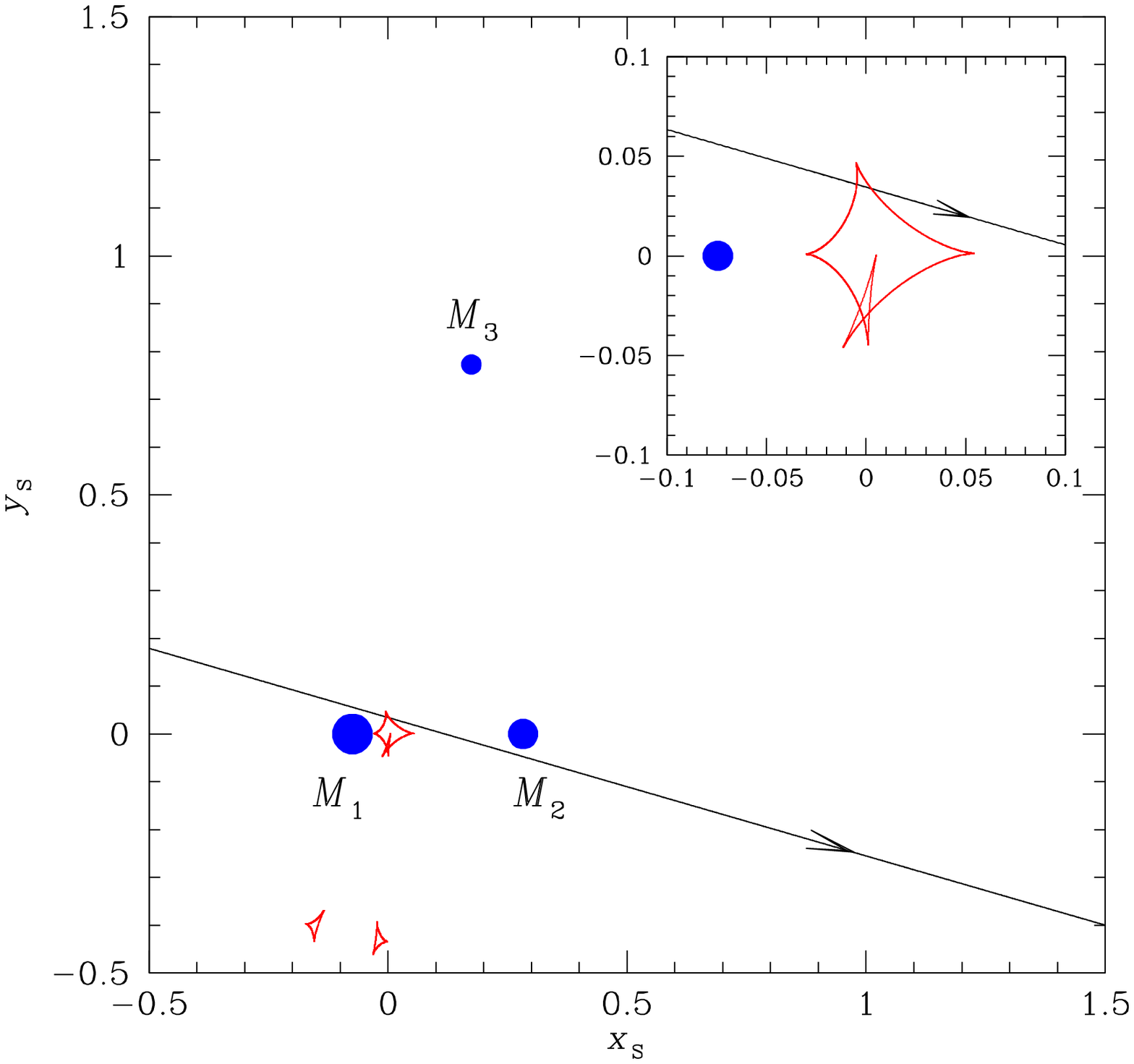}
\caption{
Lens-system configuration of the ``planet-binary'' 3L1S solution with $s_2<1.0$ 
and $s_3< 1.0$.  The blue dots marked by $M_1$, $M_2$, and $M_3$ represent the 
locations of the three lens components.  The inset shows the enlargement of the 
central caustic region.  We note that the central caustic structures of the other 
three degenerate solutions presented in Table~\ref{table:four} are similar to 
the presented one.
\smallskip
}
\label{fig:eleven}
\end{figure}

The 3L1S modeling yields three sets of solutions.  One set of solutions results 
from the starting values of $(s_2, q_2, \alpha)$ of the binary 2L1S solution and 
the other two sets result from the starting $(s_2, q_2, \alpha)$ values of the 
planetary 2L1S solution.  The solutions found based on the binary 2L1S solution 
indicates that the lens is a planetary system in a binary.  We designate these 
solutions as the ``planet-binary'' solutions.  The two sets of solutions found based 
on the planetary 2L1S solution indicate that the lens is a multiplanetary system.  
We designate these two sets of solutions as the ``multiple-planet~(I)'' and 
``multiple-planet~(II)'' solutions.  For all 3L1S solutions, the fits greatly improve 
with respect to the 2L1S and 1L2S models and the gross features of the anomalies are 
better described.  In Table~\ref{table:two}, we list the $\chi^2$ values of the 
individual solutions.  We discuss the details of the individual solutions in the 
following subsections.

\subsection{Planet-binary solution}\label{sec:five-one}

We find the ``planet-binary'' 3L1S solutions from the modeling using the initial 
values of $(s_2, q_2, \alpha)$ obtained from the ``binary'' 2L1S solution.
We find four degenerate solutions resulting from the close/wide degeneracy in the 
separations between $M_1$--$M_2$ ($s_2<1.0$ and $s_2>1.0$) and $M_1$--$M_3$ 
($s_3<1.0$ and $s_3>1.0$) pairs.  In Table~\ref{table:four}, we list the lensing 
parameters of the individual solutions along with $\chi^2$ values. It is found 
that the solutions with $s_2<1.0$ are preferred over the solutions with $s_2>1.0$ 
by $\Delta\chi^2\gtrsim 40$. For the two solutions with $s_2<1.0$, however, the 
degeneracy between the solutions with $s_3<1.0$ and $s_3>1.0$ is very severe,  
i.e., $\Delta\chi^2\sim 0.5$.

In Figure~\ref{fig:ten}, we present the model light curve (blue dashed curve) of the 
best-fit planet-binary solution (with $s_2<1.0$ and $s_3<1.0$) and the residual from 
the model.  In Figure~\ref{fig:eleven}, we also present the lens-system configuration 
corresponding to the solution. We note that the central caustic structures of the other 
degenerate solutions are similar to the presented one.  According to the planet-binary 
solutions, the source trajectory passed the upper tip of the central caustic, producing 
two caustic-crossing spikes at ${\rm HJD}^\prime\sim 8284.3$ (caustic entrance) and 
8284.5 (caustic exit), but the crossings happened  in the region of the data ending just 
before the caustic entrance and starting just after the caustic exit.  See the enlarge 
view of the crossing-crossing region of the light curve presented in the lower middle 
inset of the top panel in Figure~\ref{fig:ten}.  As a result, the value of the normalized 
source radius $\rho$ cannot be securely measured and its value is not presented in 
Table~\ref{table:four}.

\begin{figure}
\includegraphics[width=\columnwidth]{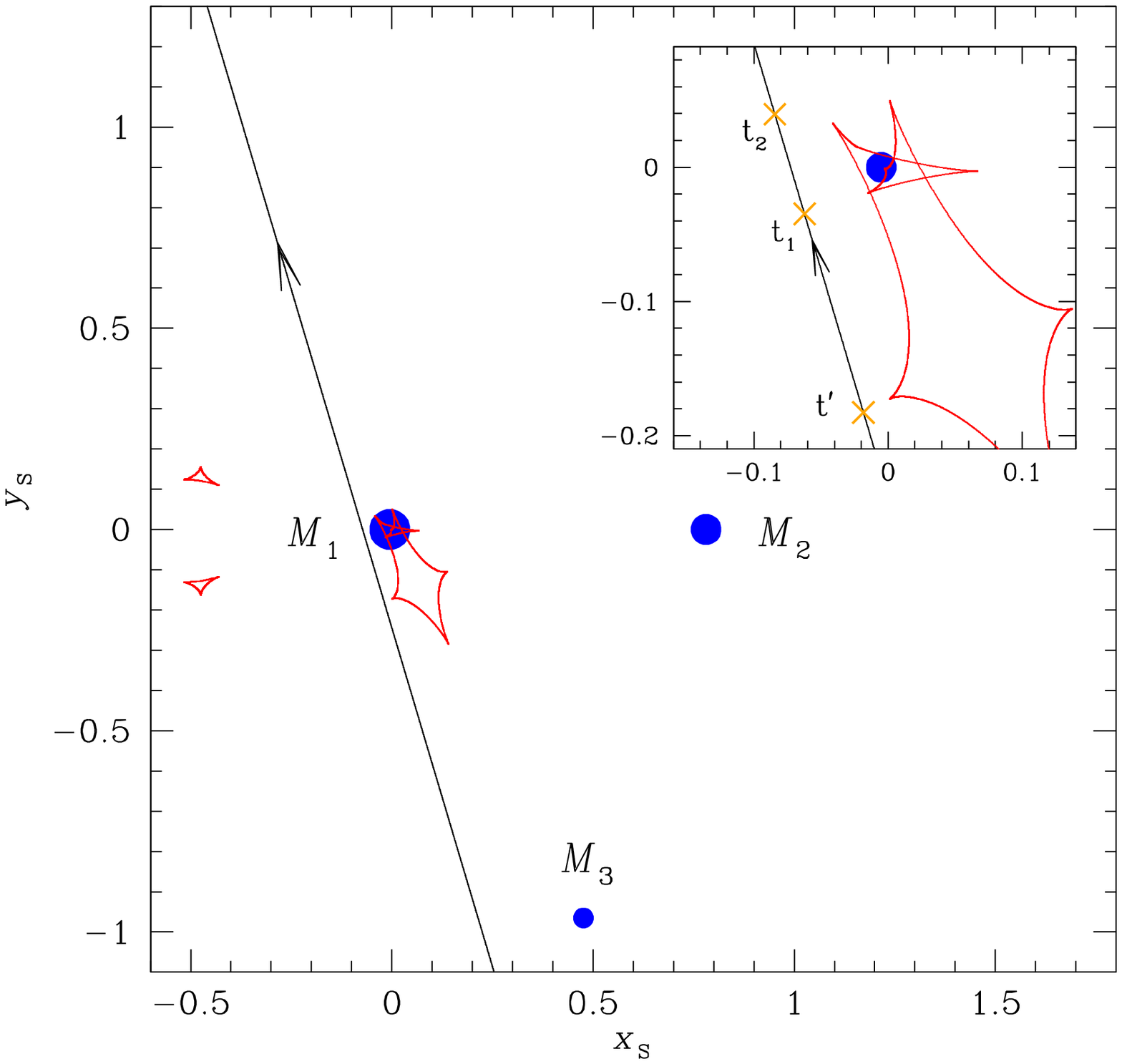}
\caption{
Lens-system configuration of the ``multiple-planet~(I)'' 3L1S solution with $s_2<1.0$.  
Notations are same as those in Fig.~\ref{fig:eleven}.  
The three ``X'' marks in the inset represent the source positions 
at the times of the cusp approaches at
$t_1=8284.6$, 
$t_2=8285.4$, and
$t^\prime=8283.0$. 
\smallskip
}
\label{fig:twelve}
\end{figure}

By comparing the lens-system configuration of the ``planet-binary'' 3L1S solution
(presented in Figure~\ref{fig:eleven}) with that of the ``binary'' 2L1S solution 
(presented in the upper panel in Figure~\ref{fig:six}), one finds that a tiny 
wedge-shape caustic appears due to the additional planetary companion with a mass 
ratio of $q_3\sim (1.7-2.8)\times 10^{-3}$.  From the comparison with the ``binary'' 
2L1S model (presented in Figure~\ref{fig:five}), it is found that the introduction 
of the third body substantially reduces the residuals of the 2L1S solution in the 
region $8284.5\lesssim {\rm HJD}^\prime \lesssim 8285.1$ and improves the fit by 
$\Delta\chi^2 \sim 645$.  We find that the fit improvement by the microlens-parallax 
and lens-orbital effects is negligible.

\begin{deluxetable}{lll}
\tablecaption{Lensing parameters of ``multiple-planet~(I)'' 3L1S solutions\label{table:five}}
\tablewidth{240pt}
\tablehead{
\multicolumn{1}{c}{Parameter}      &
\multicolumn{1}{c}{$s_2<1.0$}      & 
\multicolumn{1}{c}{$s_2>1.0$}  
}
\startdata                 
$\chi^2$                   &  7783.8                              &  7790.7                 \\
$t_0$ (${\rm HJD}^\prime$) &  8284.761 $\pm$ 0.004                &  8284.756 $\pm$ 0.004   \\
$u_0$                      &  0.070 $\pm$ 0.001                   &  0.072 $\pm$ 0.001      \\
$t_{\rm E}$ (days)         &  10.30 $\pm$ 0.11                    &  10.37 $\pm$ 0.11       \\
$s_2$                      &  0.786 $\pm$ 0.003                   &  1.193 $\pm$ 0.005      \\
$q_2$ $(10^{-3})$          &  6.47  $\pm$ 0.14                    &  6.60  $\pm$ 0.16       \\
$\alpha$ (rad)             &  4.424 $\pm$ 0.006                   &  4.424 $\pm$ 0.006      \\
$s_3$                      &  1.078 $\pm$ 0.005                   &  1.076 $\pm$ 0.004      \\
$q_3$ $(10^{-3})$          &  3.70  $\pm$ 0.15                    &  3.75  $\pm$ 0.14       \\
$\psi$ (rad)               &  5.175 $\pm$ 0.006                   &  5.174 $\pm$ 0.006      
\enddata                            
\tablecomments{${\rm HJD}^\prime={\rm HJD-2450000}$. 
\bigskip
}
\end{deluxetable}

\subsection{Multiple-planet~(I) solution}\label{sec:five-two}

The ``multiple-planet~(I)'' 3L1S solution set is one of the two sets of solutions 
obtained using the starting values of $(s_2,q_2,\alpha)$ from the ``planetary'' 
2L1S solution.  For this set of solutions, we find two degenerate solutions caused 
by the close/wide degeneracy in estimating the $M_1$--$M_2$ separation, $s_2$. The 
projected separation between $M_1$ and $M_3$ is very close to unity, $s_3\sim 1.08$, 
and thus there is no close/wide degeneracy in estimating $s_3$. It is found that the 
solution with $s_2<1.0$ is slightly preferred over the solution with $s_2>1.0$ by 
$\Delta\chi^2=6.9$.  Both companions $M_2$ and $M_3$ have planetary mass ratios of 
$q_2\sim 6.5\times 10^{-3}$ and $q_3\sim 3.7\times 10^{-3}$, respectively, indicating 
that the lens is a multiplanetary system composed of two giant planets.  In 
Table~\ref{table:five}, we list the lensing parameters for both solutions with 
$s_2<1.0$ and $s_2>1.0$.  The ``multiple-planet~(I)'' solution with $s<1.0$ provides 
a better fit than the planetary planetary 2L1S solution by $\Delta\chi^2=555.9$.

In Figure~\ref{fig:ten}, we plot the model light curve of the ``multiple-planet~(I)'' 
solution with $s_2<1.0$ (red dotted curve) and the residuals from the model.  In 
Figure~\ref{fig:twelve}, we also present the lens-system configurations corresponding 
to the solution.  It is found that the third body $M_3$ induces a resonant caustic 
in the central magnification region in addition to the central caustic induced by 
$M_2$.  We note that the solution with $s_2>1.0$ results in a similar central caustic.  
In the inset of Figure~\ref{fig:twelve}, we mark the positions of the source corresponding 
to the two major bumps at $t_1={\rm HJD}^\prime \sim 8284.6$ and $t_2\sim 8285.4$.  It 
is found the first bump at $t_1$ is produced when the source approaches the back-end 
cusp of the caustic induced by $M_2$, while the second bump at $t_2$ is produced by 
the source approach close to one of the back-end cusps of the caustic induced by $M_3$.  
We note that the source approaches another cusp induced by $M_3$ at the position marked 
by $t^\prime\sim 8283.0$ in the inset of Figure~\ref{fig:twelve}.  This approach also 
produces a bump, although the bump is weak.  See the upper left inset in the top panel 
of Figure~\ref{fig:ten}.  We note that, unlike the binary-planet solution, the 
``multiple-planet~(I)'' solution explains the anomaly without a caustic-crossing feature.  
As a result, and similarly to the binary-planet solution, the normalized source radius 
$\rho$ cannot be measured and it is not presented in Table~\ref{table:five}.  Similar 
to the case of the binary-planet solution, the higher-order effects are not important 
for the description of the observed light curve.

\begin{deluxetable*}{lcccc}
\tablecaption{Lensing parameters of ``multiple-planet~(II)'' 3L1S solutions\label{table:six}}
\tablewidth{500pt}
\tablehead{
\multicolumn{1}{c}{Parameter}                 &
\multicolumn{1}{c}{$s_2<1.0$, $s_3<1.0$}      & 
\multicolumn{1}{c}{$s_2<1.0$, $s_3>1.0$}      & 
\multicolumn{1}{c}{$s_2>1.0$, $s_3<1.0$}      & 
\multicolumn{1}{c}{$s_2>1.0$, $s_3>1.0$}  
}
\startdata                 
$\chi^2$                   &  7718.0                              &   7761.6                 &    7717.7                 &   7756.5                           \\
$t_0$ (${\rm HJD}^\prime$) &  8284.818 $\pm$ 0.005                &   8284.868 $\pm$ 0.009   &    8284.801 $\pm$ 0.005   &   8284.851 $\pm$ 0.009             \\
$u_0$                      &  0.049 $\pm$ 0.001                   &   0.047 $\pm$ 0.001      &    0.053 $\pm$ 0.001      &   0.052 $\pm$ 0.001                \\
$t_{\rm E}$ (days)         &  12.19 $\pm$ 0.14                    &   12.41 $\pm$ 0.21       &    12.42 $\pm$ 0.15       &   12.53 $\pm$ 0.21                 \\
$s_2$                      &  0.750 $\pm$ 0.005                   &   0.747 $\pm$ 0.007      &    1.281 $\pm$ 0.009      &   1.276 $\pm$ 0.011                \\
$q_2$  ($10^{-3}$)         &  9.25  $\pm$ 0.21                    &   9.73  $\pm$ 0.30       &    9.84  $\pm$ 0.26       &   10.22 $\pm$ 0.37                 \\
$\alpha$ (rad)             &  4.361 $\pm$ 0.008                   &   4.433 $\pm$ 0.008      &    4.360 $\pm$ 0.008      &   4.430 $\pm$ 0.008                \\
$s_3$                      &  0.577 $\pm$ 0.005                   &   1.954 $\pm$ 0.048      &    0.582 $\pm$ 0.005      &   1.929 $\pm$ 0.047                \\
$q_3$  ($10^{-3}$)         &  15.24 $\pm$ 0.59                    &   16.06 $\pm$ 1.70       &    15.00 $\pm$ 0.61       &   15.70 $\pm$ 1.76                 \\
$\psi$ (rad)               &  4.859 $\pm$ 0.010                   &   4.733 $\pm$ 0.011      &    4.858 $\pm$ 0.009      &   4.740 $\pm$ 0.011                \\   
$\rho$ ($10^{-3}$)         &  11.26 $\pm$ 0.72                    &   11.29 $\pm$ 0.76       &    12.13 $\pm$ 0.71       &   11.72 $\pm$ 0.75      
\enddata                            
\tablecomments{${\rm HJD}^\prime={\rm HJD-2450000}$. 
\bigskip
}
\end{deluxetable*}

\begin{figure}
\includegraphics[width=\columnwidth]{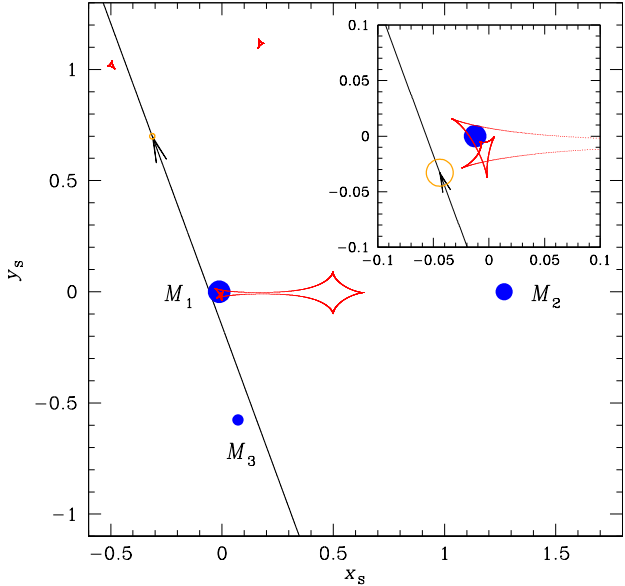}
\caption{
Lens-system configuration of the ``multiple-planet~(II)'' 3L1S solution with $s_2>1.0$
and $s_3<1.0$.  Notations are same as those in Figure~\ref{fig:eleven}.  
The small empty circle in the inset represents the source and it 
is presented to show the scaled size of the source relative to the caustic.
\smallskip
}
\label{fig:thirteen}
\end{figure}

\begin{figure}
\includegraphics[width=\columnwidth]{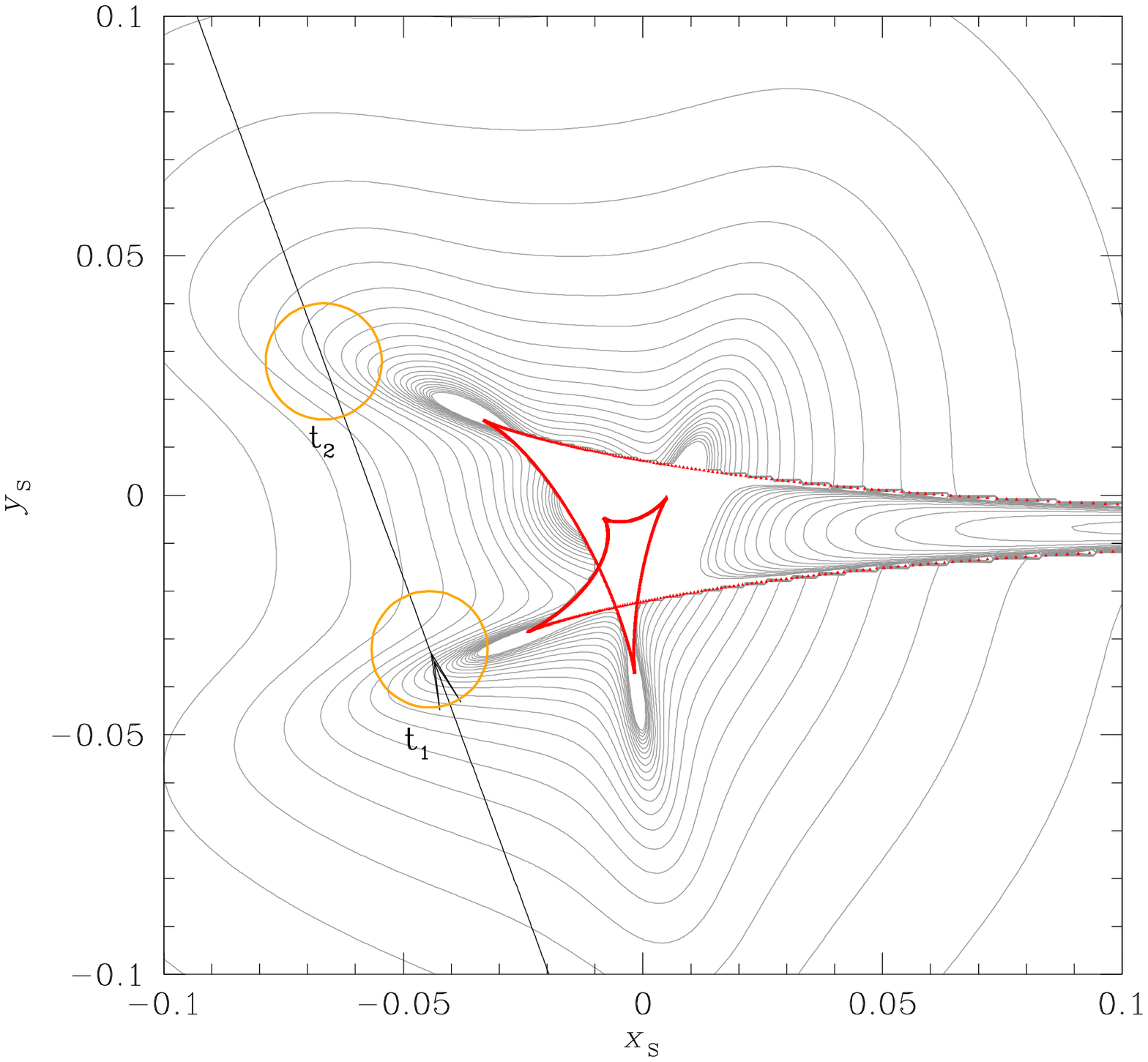}
\caption{
Magnification pattern around the central caustic of the ``multiple-planet~(II)'' solution 
with $s_2>1.0$ and $s_3<1.0$.  The innermost contour is drawn at $A=50$ and the other 
contours are drawn at the descending magnifications with a step $\Delta A=2$ from the 
center toward outward.  The two empty circles represent the source locations at the 
times of the two major bumps at $t_1=8284.6$ and $t_2=8285.4$, and the size of the circle 
is scaled to the source size.
\smallskip
}
\label{fig:fourteen}
\end{figure}

\subsection{Multiple-planet~(II) solution }\label{sec:five-three}

For given $(s_2, q_2, \alpha)$ values of the ``planetary'' 2L1S solution, we 
find another set of 3L1S solutions.  We designate these solutions as the 
``multiple-planet~(II)'' solutions.  The mass ratios between $M_1$ and $M_3$ of 
the ``multiple-planet~(II)'' solutions is $q_3\sim 15\times 10^{-3}$, which 
is considerably bigger than mass ratios of the ``multiple-planet~(I)'' solutions of 
$\sim 3.7\times 10^{-3}$.  In addition, the $M_1$--$M_3$ separations of the 
``multiple-planet~(II)'' solutions are substantially different from unity, while the 
$s_3$ values of the ``multiple-planet~(I)'' solutions are close to unity.  The best-fit 
``multiple-planet~(II)'' solution (with $s_2>1.0$ and $s_3<1.0$) yields a better fit 
to the observed data than the planetary 2L1S solution by $\Delta\chi^2=722.0$.

In Table~\ref{table:six}, we list the lensing parameters of the ``multiple-planet~(II)'' 
solutions. We find that there exist 4 solutions resulting from the close/wide 
degeneracies in both $s_2$ and $s_3$.  From the comparison of the $\chi^2$ values 
of the individual solutions, it is found that the solutions with $s_3<1.0$ is 
favored over the solutions with $s_3>1.0$ by $\Delta\chi^2\sim 40$.  However, the 
two solutions resulting from the close/wide degeneracy in $s_2$ is very severe with 
$\Delta\chi^2=0.3$.  In Figure~\ref{fig:ten}, we present the model light curve of 
the solution with $s_2>1.0$ and $s_3<1.0$, which yields the best-fit to the data, 
and the residual from the model.

\begin{figure}
\includegraphics[width=\columnwidth]{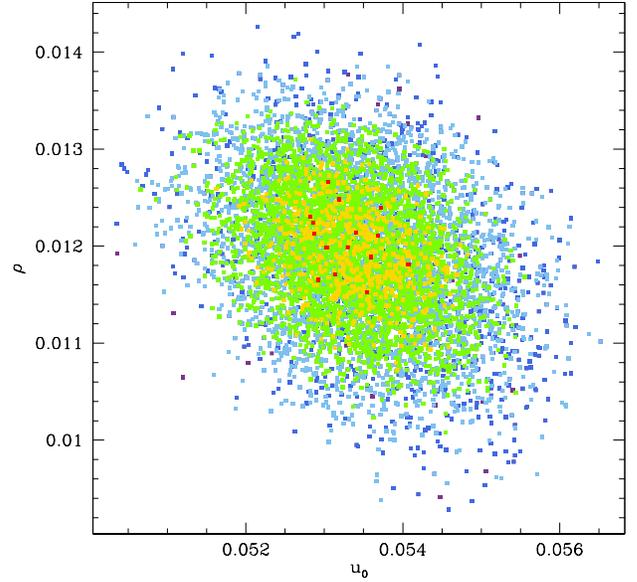}
\caption{
Distribution of points in the MCMC chain on the $u_0$--$\rho$ plane. 
The red, yellow, green, cyan, and blue colors represent points 
with $1\sigma$, $2\sigma$, $3\sigma$, $4\sigma$, and $5\sigma$, respectively.
\smallskip
}
\label{fig:fifteen}
\end{figure}

In Figure~\ref{fig:thirteen}, we present the lens-system configuration of the best-fit 
``multiple-planet~(II)'' solution.  The central caustic appears 
to be the superposition of the two sets of central caustics induced by the $M_1$--$M_2$ 
and $M_1$--$M_3$ 2-body lens pairs.  According to this solution, the bumps at $t_1$ 
and $t_2$ are produced by the successive approaches of the source close to the 
back-end cusps of the central caustic produced by the $M_1$--$M_2$ binary pair.  
However, the cusp of the first source approach is deformed by $M_3$ 
and thus the central caustic is different from that of the central caustic of the 
$M_1$--$M_2$ binary.  We note that the patterns of central caustics for the other 
degenerate solutions are similar to the presented one.

For the ``multiple-planet~(II)'' solutions, the source does not cross the caustic, 
but finite-source effects are securely detected unlike the previous two sets of 
3L1S solutions.  To investigate the reason for this, we construct the magnification 
pattern around the central caustic.  Figure~\ref{fig:fourteen} shows the constructed 
magnification contour map, in which the innermost contour is drawn at $A=50$ and 
the other contours are drawn at the descending magnifications with a step $\Delta A=2$ 
from the center toward outward.  The line with an arrow represents the source trajectory 
and the two orange circles on the trajectory represent the source locations at the times 
of the first ($t_1$) and second ($t_2$) caustic approaches, respectively, and the size 
of the circle is scaled to the source size.  We find that the magnification on the 
surface of the source varies substantially and this results in deviation of the light 
curve from the point-source light curve during the times around both bumps.  From the 
deviation of the light curve, the normalized source radius is securely measured.  
Although there is some variation depending on the solution, the measured normalized 
source radius is $\rho\sim 0.012$.  In Figure~\ref{fig:fifteen}, we present the distribution 
of points in the MCMC chain on the $u_0$--$\rho$ plane for the best-fit 
``multiple-planet~(II)'' solution.  Higher-order effects are not detected for the 
solution.

\subsection{Comparison of Models}\label{sec:five-four}
\begin{figure}
\includegraphics[width=\columnwidth]{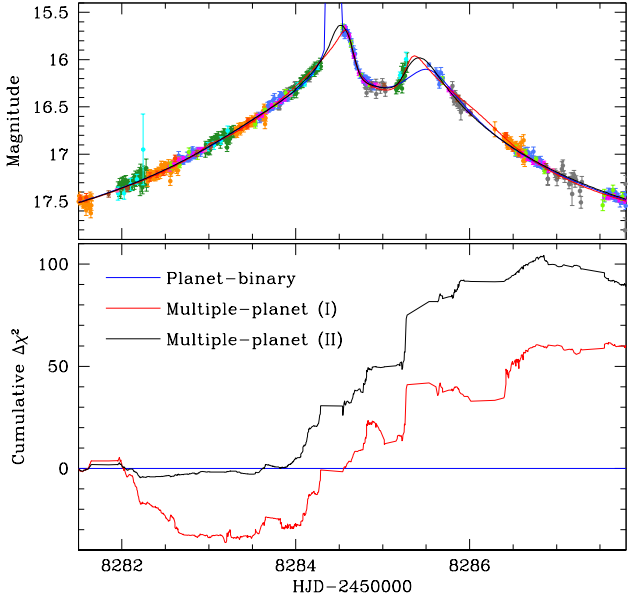}
\caption{
Cumulative distribution of 
$\Delta\chi^2=\chi^2_{\rm p-b}-\chi^2_{\rm m-p}$ between the planet-binary and 
``multiple-planet'' 3L1S solutions for the data in the regions of the anomaly.  Here 
the subscripts ``p-b'' and ``m-p'' denote planet-binary and ``multiple-planet'' solutions, 
respectively.  The light curve in the upper panel is presented to show the regions of 
$\chi^2$ difference. 
\smallskip
}
\label{fig:sixteen}
\end{figure}

Because the overall features of the observed central perturbation are described 
by three sets of 3L1S solutions, we closely investigate the individual solutions. 
For this, we construct a cumulative distribution of 
$\Delta\chi^2=\chi^2_{\rm p-b}-\chi^2_{\rm m-p}$ for the data around the region 
of the anomaly. Here the subscripts ``p-b'' and ``m-p'' denote ``planet-binary''
and ``multiple-planet'' solutions, respectively.  In Figure~\ref{fig:sixteen}, we 
present the constructed cumulative $\Delta\chi^2$ distributions.

From the cumulative $\Delta\chi^2$ distributions, together with the residuals of 
the three solutions presented in Figure~\ref{fig:ten}, we find that the 
``multiple-planet~(II)'' solution provides a better fit over the other two 
solutions. Compared to the ``planet-binary'' solution, the ``multiple-planet~(II)'' 
solution better explains the rising part of the light curve in the region of the 
anomaly at ${\rm HJD}^\prime \sim 8285.2$, resulting in a better fit by 
$\Delta\chi^2=107.7$ over the ``planet-binary'' solution.  The ``planet-binary'' 
solution is additionally disfavored by Occam's razor because the major features 
occur only during the gap in the data.  Compared to the ``multiple-planet~(I)'' 
solution, the ``multiple-planet~(II)'' solution better describes the light curve 
in the region $8282.0\lesssim {\rm HJD}^\prime\lesssim 8284.1$, resulting in a better 
fit by $\Delta\chi^2=65.9$.  We note that according to the ``multiple-planet~(I)'' 
solution, the source approached a cusp of the caustic induced by $M_3$ at around 
$t^\prime\sim 8283.0$, producing a weak bump.  However, no such a bump is expected 
according to the ``multiple-planet~(II)'' solution, and this makes the solution 
better fit the data over the ``multiple-planet~(I)'' solution.  Considering the 
better description of the detailed structures of the lensing light curve over the 
other models, we conclude that the ``multiple-planet~(II)'' solution provides the 
most plausible model of the observed data.

\section{Source Star}\label{sec:six}

Characterizing a source star in microlensing is important in order to estimate 
the angular Einstein radius, $\thetae$, that is related to the angular source 
radius $\theta_*$ by 
\begin{equation} 
\thetae={\theta_* \over \rho}.  
\label{eq2}
\end{equation} 
For OGLE-2018-BLG-1011, the normalized source radius is securely measured 
despite that the source does not cross the caustic.  Then, one needs to estimate 
the angular source radius to estimate the angular Einstein radius.

\begin{figure}
\includegraphics[width=\columnwidth]{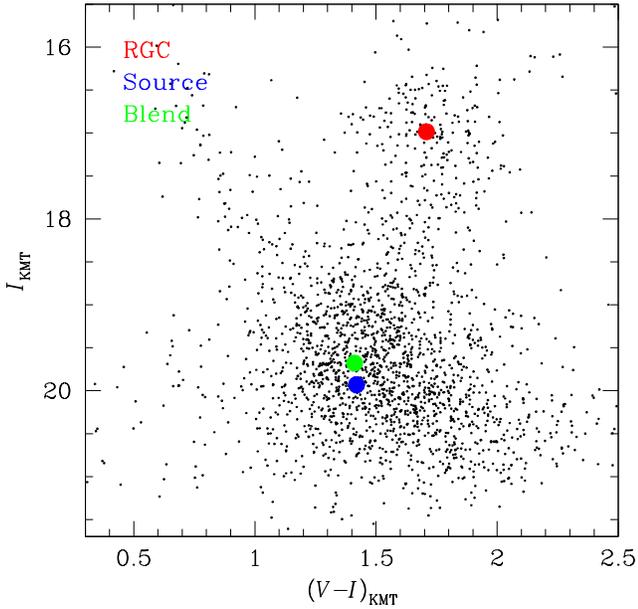}
\caption{
Position of the source with respect to the centroid of red giant clump (RGC) 
in the instrumental color-magnitude diagram of stars around the source.  
Also marked is the position of the blend.  
\bigskip
}
\label{fig:seventeen}
\end{figure}

For the estimation of the angular source radius, we first estimate the de-reddened 
color $(V-I)_{\rm S,0}$  and brightness $I_{\rm S,0}$ of the source star.  For this 
estimation, we use the \citet{Yoo2004} method, in which the color and magnitude are 
calibrated using the centroid of the red giant clump (RGC) in the color-magnitude 
diagram (CMD) as a reference.  In Figure~\ref{fig:seventeen}, we present the location 
of the source in the CMD of stars within $\sim 20^\prime$ around the source star.  
The CMD is constructed based on the pyDIA photometry of the KMTC $I$- and $V$-band data sets.
The instrumental color and brightness of the source are 
$(V-I,I)_{\rm S}=(1.42\pm 0.01, 19.93\pm 0.01)$.  
The offsets in color and magnitude of the source with respect to the RGC centroid, 
which is located at 
$(V-I,I)_{\rm RGC}=(1.71, 16.99)$, are $\Delta (V-I,I)_{\rm S}=(-0.29, 2.95)$.  
With the known de-reddened values of the RGC 
centroid of $(V-I,I)_{\rm RGC,0}=(1.06, 14.38)$ \citep{Bensby2011, Nataf2013}, then 
the de-reddened color and brightness of the source star are estimated as
$(V-I,I)_{\rm S,0}=(V-I,I)_{\rm S}+\Delta (V-I,I)_{\rm S}=(0.77\pm 0.01, 17.33\pm 0.01)$.  
These values indicate that the source is a late G-type turn-off star.  In 
Figure~\ref{fig:seventeen}, we also mark the position of the blend in the CMD. It is 
found that the blend has a color and a brightness that are similar to those of the 
source.  The brightness of the blend is $I_{\rm b}=19.04\pm 0.01$ as measured in the 
OGLE photometry system, which is approximately calibrated.

\begin{deluxetable}{ll}
\tablecaption{Source, blend, Einstein radius, and proper motion\label{table:seven}}
\tablewidth{240pt}
\tablehead{
\multicolumn{1}{c}{Quantity}      &
\multicolumn{1}{c}{Value}  
}
\startdata                 
$(V-I)_0$              &   $0.77 \pm 0.01$         \\
$I_0$                  &   $17.33\pm 0.01$         \\
$I_{\rm b}$            &   $19.04\pm 0.01$         \\
$\theta_*$ ($\mu$as)   &   $1.98 \pm 0.08$         \\
$\thetae$  (mas)       &   $0.09 \pm 0.01$         \\
$\mu$ (mas~yr$^{-1}$)  &   $2.81 \pm 0.22$          

\enddata                            
\tablecomments{
The value $I_{\rm b}$ presents the $I$-band magnitude of the blend.
\smallskip
}
\end{deluxetable}

With the known de-reddened color and brightness, the angular source radius 
is estimated first by converting the $V-I$ color into the $V-K$ color using the 
color-color relation \citep{Bessell1988} and second using the $(V-K)/\theta_*$ 
relation  \citep{Kervella2004}.  This procedure yields the angular source radius of 
\begin{equation}
\theta_* = 1.98 \pm 0.08~\mu{\rm as}.
\label{eq3}
\end{equation}
We note that two major factors affect the precision of the estimated angular source 
radius.  The first is the uncertainty of the de-reddened color, $\sim 0.05$~mag, and 
the other is the uncertainty in the position of RGC, $\sim 0.1$ mag.  The uncertainty 
of $\theta_*$ is estimated by considering the combined uncertainty, which is $\sim 7\%$, 
of these two factors \citep{Gould2014a}.  
With the measured $\theta_*$ together with $\rho$, the angular Einstein radius is 
estimated as 
\begin{equation}
\thetae = {\theta_*\over \rho } = 0.09 \pm 0.01~{\rm mas}. 
\label{eq4}
\end{equation}
With the angular Einstein radius combined with the event timescale,
the relative lens-source proper motion is estimated as 
\begin{equation}
\mu = {\thetae\over t_{\rm E}} = 2.81 \pm 0.22~{\rm mas}. 
\label{eq5}
\end{equation}
In Table~\ref{table:seven}, we summarize the colors and magnitudes of the source and 
blend and the estimated angular source radius, Einstein radius, and the relative 
lens-source proper motion.

\begin{figure}
\includegraphics[width=\columnwidth]{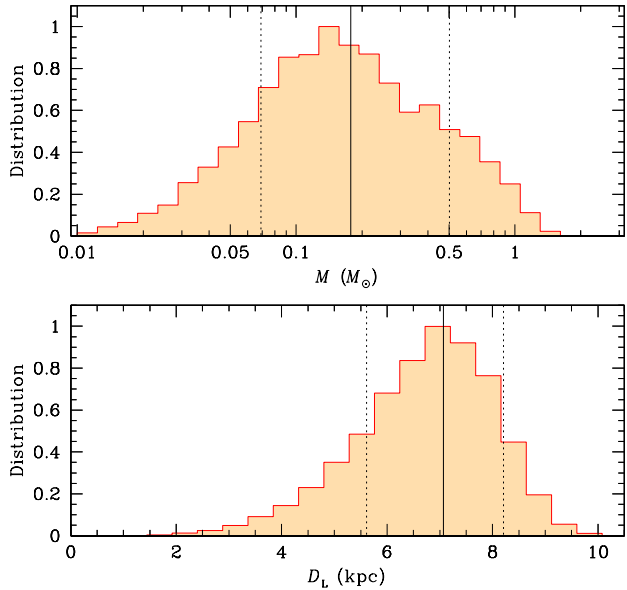}
\caption{
The distributions of the lens mass (upper panel) and distance (lower panel) estimated 
by Bayesian analysis.  The solid vertical line for each distribution indicates the 
median value, and the two dotted vertical lines represent the $1\sigma$ range of the 
estimated value, i.e., 16\% and 84\% of the distribution.   
\smallskip
}
\label{fig:eighteen}
\end{figure}

For an independent constraint on the source star distance, we check the source 
information in the list of {\it Gaia} data release 2 \citep[{\it Gaia} DR2:][]{Gaia2018}.  
However, there is no information of the absolute parallax and proper motion of the source 
because the source, with a $G$-band magnitude of $G=19.76$, is fainter than the {\it Gaia} 
limit of $G\sim 18$.  As a result, it is difficult to constraint the source distance from 
the {\it Gaia} data.

\section{Physical Parameters}\label{sec:seven}

For the unique determination of the mass, $M$, and distance to the lens, $D_{\rm L}$, 
one needs to measure both the angular Einstein radius $\thetae$ and the microlens 
parallax $\pie$, which are related to $M$ and $D_{\rm L}$ by the relations
\begin{equation}
M={\thetae\over \kappa\pie};\qquad
D_{\rm L}={{\rm au}\over \pie\thetae +\pi_{\rm S}},
\label{eq6}
\end{equation}
where $\kappa=4G/(c^2{\rm au})$, $\pi_{\rm S}={\rm au}/D_{\rm S}$, and $D_{\rm S}$ is 
the source distance, which is $\sim 8$ kpc for a source star located in the bulge. 
For OGLE-2018-BLG-1011, the angular Einstein radius is measured, but the microlens
parallax is not measured. We, therefore, estimate the physical lens parameters by 
conducting Bayesian analysis with the constraints of the measured $t_{\rm E}$ and 
$\thetae$.

A microlensing Bayesian analysis requires prior models of the lens mass function 
and the physical and dynamical distributions of the Galaxy. We adopt the mass 
function of \citet{Chabrier2003} for stars and that of \citet{Gould2000} for stellar 
remnants. For the physical and dynamical distributions, we adopt \citet{Han2003} and 
\citet{Han1995} models, respectively. For more details of these models, see 
section~5 of \citet{Han2018}. Based on these priors, we conduct a Monte Carlo simulation 
and produce $2\times 10^6$ microlensing events. We then construct the distributions $M$ 
and $D_{\rm L}$ for events that have timescales and angular Einstein radii within the 
ranges of the measured $t_{\rm E}$ and $\thetae$. We estimate $M$ and $D_{\rm L}$ as 
the median values of the distributions and their lower and upper limits are estimated 
as the 16\% and 84\% of the distributions, respectively.

In the Bayesian analysis, we also impose the constraint of the lens brightness so 
that the lens cannot be brighter than the measured blend brightness of 
$I_{\rm b}\sim 19.0$.  The lens brightness is computed based on the mass, distance, 
and extinction. We assume that the extinction linearly increases with distance until 
it becomes $A_I\sim 1.68$ at $D_{\rm S}$, which is the measured value of the extinction 
toward the field.  We find that this constraint has little effect on the lens mass and 
distance distributions. 
This is because lenses, in most cases, are much fainter than the blend.
For the same reason, the result would not change with the choice of 
different extinction model along the line of sight.

\begin{deluxetable}{ll}
\tablecaption{Physical lens parameters\label{table:eight}}
\tablewidth{240pt}
\tablehead{
\multicolumn{1}{c}{Parameter}                      &
\multicolumn{1}{c}{Value}  
}
\startdata                 
$M_1$ ($M_\odot$)    &   $0.18^{+0.33}_{-0.10}$                      \\
$M_2$ ($M_{\rm J}$)  &   $1.8^{+3.4}_{-1.1}$                         \\
$M_3$ ($M_{\rm J}$)  &   $2.8^{+5.1}_{-1.7}$                         \\
$D_{\rm L}$ (kpc)    &   $7.1^{+1.1}_{-1.5}$                         \\
$a_{\perp,2}$ (au)   &   $1.8^{+2.1}_{-1.5}$  ($0.8^{+0.9}_{-0.6}$)  \\
$a_{\perp,3}$ (au)   &   $0.8^{+0.9}_{-0.6}$                  
\enddata                            
\tablecomments{
The $a_{\perp,2}$ value in the parenthesis is for the solution with $s_2>1.0$.
}
\end{deluxetable}

In Figure~\ref{fig:eighteen}, we present the distributions of the lens mass and 
the distance obtained from the Bayesian analysis.  In Table~\ref{table:eight}, 
we list the estimated values of the individual lens components, $M_1$, $M_2$, and 
$M_3$, the distance to the lens, $D_{\rm L}$, and the projected separations of the 
planets from the host, $a_{\perp, 2}$ and $a_{\perp, 3}$.  It is found that the 
lens is a multiple-planet system composed of two giant planets with masses
\begin{equation}
M_2=1.8^{+3.4}_{-1.1}~M_{\rm J}
\label{eq7}
\end{equation}
and
\begin{equation}
M_3=2.8^{+5.1}_{-1.7}~M_{\rm J}
\label{eq8}
\end{equation}
around a host star with a mass
\begin{equation}
M_1=0.18^{+0.33}_{-0.10}~M_\odot.
\label{eq9}
\end{equation}
The distance to the lens is estimated as
\begin{equation}
D_{\rm L}=7.1^{+1.1}_{-1.5}~{\rm kpc}.
\label{eq10}
\end{equation}
The estimated distance indicates that the lens is the farthest system among the 
known multiplanetary systems. We note that the previous multiplanetary system with 
the farthest distance is OGLE-2012-BLG-0026L that is located at $\sim 4.0\pm 0.3$~kpc 
away \citep{Han2013, Beaulieu2016}.  The projected separations of the planets from the 
host are
\begin{equation}
a_{\perp,2}=1.8^{+2.1}_{-1.5}~\left(0.8^{+0.9}_{-0.6}\right)~{\rm au}
\label{eq11}
\end{equation}
and
\begin{equation}
a_{\perp,3}=0.8^{+0.9}_{-0.6}~{\rm au}.
\label{eq12}
\end{equation}
We note that the value of $a_{\perp,2}$ in the parenthesis of equation~(\ref{eq11}) 
is for the solution with $s_2<1.0$.  The snow line of the planetary system is 
$a_{\rm sl}\sim 2.7~{\rm au}~(M/M_\odot)\sim 0.5~{\rm au}$, and thus both planets 
are located beyond the snow line of the host similar to the other cases of the 
multiplanetary systems found by microlensing.

\section{Discussion and Conclusion}\label{sec:eight}

We investigated the microlensing event OGLE-2018-BLG-1011, for which the light 
curve exhibited an anomaly around the peak. We found that it was not possible to 
reasonably explain the anomaly with the binary-lens or binary-source interpretations 
and its description required the introduction of an additional lens component.  The 
3L1S modeling resulted in three sets of solutions, in which one set of solutions 
indicated that the lens was a planetary system in a binary, while the other two sets 
of solutions implied that the lens was a multiplanetary system.  By investigating 
the fits of the individual models to the detailed structure of the lensing light 
curve, we found that the multiple-planet solutions with planet-to-host mass ratios 
$\sim 9.5\times 10^{-3}$ and $15\times 10^{-3}$ were favored over the other solutions.  
From the Bayesian analysis for the best-fit solution, it was found that the lens is a 
multiple planetary system composed of giant planets with masses $\sim 1.8~M_{\rm J}$ 
and $\sim 2.8~M_{\rm J}$ orbiting a bulge star with a mass $\sim 0.18~M_\odot$ located 
at a distance of $\sim 7.1~{\rm  kpc}$.  The projected separations of the planets from 
the host were $a_{\perp,2}\sim 1.8~{\rm au}$ (or $\sim 0.8~{\rm au}$) and 
$a_{\perp,3}\sim 0.8~{\rm au}$, where the values of $a_{\perp,2}$ denoted 
with and without parentheses were the separations corresponding to the two degenerate 
solutions with close and wide separations. Therefore, both planets were located beyond 
the snow line of the host similar to the other four multiplanetary systems previously 
found by microlensing.

\acknowledgments
Work by CH was supported by the grant (2017R1A4A1015178) of National Research Foundation of Korea.
Work by AG was supported by US NSF grant AST-1516842.
AG received support from the European Research Council under the European Union's
Seventh Framework Programme (FP 7) ERC Grant Agreement n.~[321035].
The MOA project is supported by JSPS KAKENHI Grant Number JSPS24253004,
JSPS26247023, JSPS23340064, JSPS15H00781, and JP16H06287.
YM acknowledges the support by the grant JP14002006.
DPB, AB, and CR were supported by NASA through grant NASA-80NSSC18K0274. 
The work by CR was supported by an appointment to the NASA Postdoctoral Program at the Goddard 
Space Flight Center, administered by USRA through a contract with NASA. NJR is a Royal Society 
of New Zealand Rutherford Discovery Fellow.
The OGLE project has received funding from the National Science Centre, Poland, grant
MAESTRO 2014/14/A/ST9/00121 to AU.
This research has made use of the KMTNet system operated by the Korea
Astronomy and Space Science Institute (KASI) and the data were obtained at
three host sites of CTIO in Chile, SAAO in South Africa, and SSO in
Australia.
This research uses data obtained through the Telescope Access Program (TAP), which 
has been funded by the National Astronomical Observatories of China, the Chinese 
Academy of Sciences, and the Special Fund for Astronomy from the Ministry of Finance. 
WZ, WT, and SM acknowledges support by the National Science Foundation of China 
(Grant 11821303 and 11761131004). Work by WZ, and PF was supported by 
Canada-France-Hawaii Telescope (CFHT).
UKIRT is currently owned by the University of Hawaii (UH) and operated by the 
UH Institute for Astronomy; operations are enabled through the cooperation of the 
East Asian Observatory. The collection of the 2018 data reported here was supported 
by NASA grant NNG16PJ32C and JPL proposal \#18-NUP2018-0016.
We acknowledge the high-speed internet service (KREONET)
provided by Korea Institute of Science and Technology Information (KISTI).

\end{document}